\input harvmac
\noblackbox

\def\Im{{\rm Im}}
\def\Re{{\rm Re}}
\def\cof{{\rm cof}}

\def\axx{(\alpha_x \cdot \alpha_x)}
\def\axy{(\alpha_x \cdot \alpha_y)}
\def\ayy{(\alpha_y \cdot \alpha_y)}
\def\bxx{(\beta_x \cdot \beta_x)}
\def\bxy{(\beta_x \cdot \beta_y)}
\def\byy{(\beta_y \cdot \beta_y)}
\def\gxx{(\alpha_x \cdot \beta_x)}
\def\gyy{(\alpha_y \cdot \beta_y)}
\def\gxy{\alpha_x \cdot \beta_y}
\def\gyx{\alpha_y \cdot \beta_x}

\def\ax{\alpha_x}
\def\bx{\beta_x}
\def\ay{\alpha_y}
\def\by{\beta_y}

\lref\oldflux{S. Gukov, C. Vafa and E. Witten,
``CFTs from Calabi-Yau Fourfolds,'' Nucl. Phys. {\bf B584}, 69
(2000) [arXiv:hep-th/9906070]}

\lref\DRS{K. Dasgupta, G. Rajesh and S. Sethi, ``M-theory, Orientifolds
and G-flux,'' JHEP {\bf 9908}, 023 (1999) [arXiv:hep-th/9908088].}

\lref\TV{T. Taylor and C. Vafa, ``RR flux on Calabi-Yau and Partial 
Supersymmetry Breading,'' Phys. Lett. {\bf B474}, 130 (2000)
[arXiv:hep-th/9912152].
}

\lref\curio{G. Curio, A. Klemm, D. Lust and S. Theisen, ``On the
Vacuum Structure of Type II String Compactifications on Calabi-Yau
Spaces with H Fluxes,'' Nucl. Phys. {\bf B609}, 3 (2001) [arXiv:
hep-th/0012213].}

\lref\louis{M. Haack and J. Louis, ``M-theory Compactified on a
Calabi-Yau Fourfold with Background Flux,'' Phys. Lett. {\bf B507},
296 (2001) [arXiv:hep-th/0103068].}
 
\lref\GKP{
S.~B.~Giddings, S.~Kachru and J.~Polchinski,
``Hierarchies from fluxes in string compactifications,''
[arXiv:hep-th/0105097].
}

\lref\EvaFix{E. Silverstein, ``(A)dS Backgrounds from Asymmetric
Orientifolds,'' [arXiv:hep-th/0106209].}

\lref\Beckerst{
K. Becker and M. Becker, ``Supersymmetry Breaking, M-theory
and Fluxes,'' JHEP {\bf 0107}, 038 (2001) [arXiv:hep-th/0107044].}

\lref\kst{S. Kachru, M. B. Schulz and S. P. Trivedi,
``Moduli Stabilization From Fluxes in a Simple IIB Orientifold,"
[arXiv:hep-th/0201028].}
  
\lref\kstt{S. Kachru, M. B. Schulz, P. K. Tripathy and S. P. Trivedi,
``New Supersymmetric String Compactifications,'' [arXiv:hep-th/0211182].}

\lref\joefrey{A. Frey and J. Polchinski, ``${\cal N}=3$ Warped
Compactifications,'' [arXiv:hep-th/0201029].} 

\lref\acfh{R. Argurio, V. Campos, G. Ferretti and R. Heise, ``Freezing
of Moduli with Fluxes in Three-Dimensions,'' Nucl. Phys. {\bf B640}, 351
(2002) [arXiv: hep-th/0205295].
} 

\lref\Beckers{K. Becker and M. Becker, ``M-theory on Eight Manifolds,''
Nucl. Phys. {\bf B477}, 155 (1996) [arXiv:hep-th/9605053].} 

\lref\GSS{B. Greene, K. Schalm and G. Shiu, ``Warped Compactifications in
M and F-theory,'' Nucl. Phys. {\bf B584}, 480 (2000)
[arXiv:hep-th/0004103].}
  
\lref\Herman{H. Verlinde, ``Holography and Compactification,''
Nucl. Phys. {\bf B580}, 264 (2000) [arXiv:hep-th/9906182]\semi
C. Chan, P. Paul and H. Verlinde, ``A Note on Warped String
Compactification,'' Nucl. Phys. {\bf B581}, 156 (2000)
[arXiv:hep-th/0003236].}

\lref\Mayr{P. Mayr, ``On Supersymmetry Breaking in String Theory
and its Realization in Brane Worlds,'' Nucl. Phys. {\bf B593}, 99
(2001) [arXiv:hep-th/0003198]\semi
P. Mayr, ``Stringy Brane Worlds and Exponential Hierarchies,''
JHEP {\bf 0011}, 013 (2000), [arXiv:hep-th/0006204].}

\lref\PolchinskiSM{
J.~Polchinski and A.~Strominger,
``New Vacua for Type II String Theory,''
Phys.\ Lett.\ B {\bf 388}, 736 (1996)
[arXiv:hep-th/9510227].
} 

\lref\MichelsonPN{
J.~Michelson,
``Compactifications of type IIB strings to four dimensions
with  non-trivial classical potential,''
Nucl.\ Phys.\ B {\bf 495}, 127 (1997)
[arXiv:hep-th/9610151].
} 

\lref\deBoerPX{
J.~de Boer, R.~Dijkgraaf, K.~Hori, A.~Keurentjes, J.~Morgan,
D.~R.~Morrison and S.~Sethi,
``Triples, fluxes, and strings,''
Adv.\ Theor.\ Math.\ Phys.\  {\bf 4}, 995 (2002)
[arXiv:hep-th/0103170].
}

\lref\Atish{
A.~Dabholkar and C.~Hull,
``Duality Twists, Orbifolds, and Fluxes,''
[arXiv:hep-th/0210209].
}

\lref\klst{S. Kachru, X. Liu, M. Schulz and S. P. Trivedi, ``Supersymmetry
Changing Bubbles in String Theory,'' [arXiv:hep-th/0205108].}

\lref\keshav{K. Becker and K. Dasgupta, ``Heterotic Strings with
Torsion,'' [arXiv:hep-th/0209077]
} 

\lref\janlouis{S. Gurrieri, J. Louis, A. Micu and D. Waldram, ``Mirror
Symmetry in Generalized Calabi-Yau Compactifications,''
[arXiv:hep-th/0211102].}
 
\lref\dlust{G. Cardoso, G. Curio, G. Dall'Agata, D. L\"ust, P. Manousselis
and G. Zoupanos, ``Non-K\"ahler String Backgrounds and their Five Torsion
Classes,'' [arXiv:hep-th/0211118].}

\lref\kstt{S. Kachru, M. B. Schulz, P. K. Tripathy and S. P. Trivedi,
``New Supersymmetric String Compactifications,'' [arXiv:hep-th/0211182].} 

\lref\micu{
S.~Gurrieri and A.~Micu,
 ``Type IIB theory on half-flat manifolds,'' [arXiv:hep-th/0212278].
}

\lref\GP{ E.~Goldstein and S.~Prokushkin, ``Geometric Model
for Complex non-Kaehler Manifolds with SU(3) Structure,''
 [arXiv:hep-th/0212307].
}

\lref\aspin{P. S. Aspinwall, ``K3 Surfaces and String Duality,''
[arXiv:hep-th/9611137].
} 

\lref\djm{K. Dasgupta, D. P. Jatkar and S. Mukhi, ``Gravitational Couplings
and $Z_2$ Orientifolds,'' Nucl. Phys. {\bf B523}, 465 (1998)
[arXiv:hep-th/9707224] \semi
B. Stefanski, ``Gravitational Couplings of D-branes and O-planes,''
Nucl. Phys. {\bf B548}, 275 (1999) 
[arXiv:hep-th/9812088] \semi
J. F. Morales, C. A. Scrucca and M. Serone, ``Anomalous Couplings for
D-branes and O-planes,''
Nucl. Phys. {\bf B552}, 291 (1999)
[arXiv:hep-th/9812071]. 
} 

\lref\rBUSH{T. Buscher, {\it ``Quantum Corrections and Extended
Supersymmetry in New Sigma Models''}, Phys. Lett. {\bf B159}
(1985) 127; {\it ``A Symmetry Of The String Background Field
Equations''}, Phys. Lett. {\bf B194} (1987) 59; {\it ``Path
Integral Derivation of Quantum Duality in Nonlinear Sigma
Models''}, Phys. Lett. {\bf B201} (1988) 466.}

\lref\HassanBV{
S.~F.~Hassan,
``T-duality, space-time spinors and R-R fields in curved backgrounds,''
Nucl.\ Phys.\ B {\bf 568}, 145 (2000)
[arXiv:hep-th/9907152].
} 

\lref\BHO{
E.~Bergshoeff, C.~Hull and T.~Ortin,
``Duality in the type-II superstring effective action,''
Nucl.\ Phys.\ B {\bf 451}, 547 (1995).
}  

\lref\haridass{
B.~de Wit, D.~J.~Smit and N.~D.~Hari Dass,
``Residual Supersymmetry Of Compactified D = 10 Supergravity'',
Nucl.\ Phys.\ {\bf B283} (1987) 165 (1987).}

\lref\andytors{
{A. ~Strominger,  ``Superstrings With Torsion'',
Nucl.\ Phys.\ {\bf B274} (1986) 253.}
}

\lref\granapolch{M.~Grana and J.~Polchinski,
``Supersymmetric three-form flux perturbations on AdS(5),''
Phys.\ Rev.\ D {\bf 63}, 026001 (2001)
[arXiv:hep-th/0009211].
}

\lref\RS{L. Randall and R. Sundrum, ``A Large Mass Hierarchy
from a Small Extra Dimension,'' Phys. Rev. Lett. {\bf 83}, 3370
(1999) [arXiv:hep-ph/9905221].}

\lref\ferrara{
S. Ferrara and  M. Porrati,
``N=1 No Scale Supergravity From IIB Orientifolds,''
Phys. Lett. {\bf 545}, 411 (2002)
[arXiv:hep-th/0207135].
}

\lref\dauria{
R. D'Auria, S. Ferrara and S. Vaula,
``N=4 Gauged Supergravity And a IIB Orientifold With Fluxes,''
New J. Phys. {\bf 4}, 71 (2002)
[arXiv:hep-th/0206241].
}

\lref\lledo{
L. Andrianopoli, R. D'Auria, S. Ferrara and  M. A. Lledo,
``Gauging of Flat Groups in Four Dimensional Supergravity,''
JHEP {\bf 0207} 010 (2002)
[arXiv:hep-th/0203206].
}

\lref\add{N.  Arkani-Hamed, S. Dimopoulos and  G.R. Dvali,
``The Hierarchy Problem and New Dimensions at a Millimeter",
Phys.Lett. {\bf 429}, 263 (1998) 
[arXiv: hep-ph/9803315].
}

\lref\mr{N. Arkani-Hamed, S. Dimapoulos and J. March-Russell,
``Stabilisation of Submillimeter Dimensions: The New Guise of the Hierarchy Problem,"
 Phys.Rev.D {\bf 63}, 064020,2001 
[arXiv: hep-th/9809124].
}  

\lref\bbdg{K. Becker, M. Becker, K. Dasgupta and P. Green, 
``Compactifications of Heterotic Theory on Non-K\"ahler Complex Manifolds: I", 
[arXiv: hep-th/0301161]. }

\lref\dagata{G. Dall'Agata, ``Type IIB Supergravity Compactified on a Calabi-Yau
Manifold with H Fluxes," 
JHEP {\bf 0111} 005 (2001)
[arXiv: hep-th/0107264].
}

\Title{\vbox{\baselineskip12pt
\hbox{hep-th/0301139}
\hbox{TIFR/TH/02-34}
}}
{\vbox{\centerline{Compactification with Flux on K3 and Tori}}}

\centerline{Prasanta K. Tripathy \footnote{$^1$}{prasanta@theory.tifr.res.in}
and
Sandip P. Trivedi \footnote{$^2$}{sandip@tifr.res.in}}
\medskip
\centerline{\it Tata Institute of Fundamental Research}
\centerline{\it Homi Bhabha Road, Mumbai 400 005, INDIA}

\vskip .15in
We study compactifications of Type IIB string theory on  a $K3 \times T^2/Z_2$ orientifold
in the presence of  RR and NS flux. We find the most general supersymmetry preserving,
Poincare invariant, vacua in this model.
 All the complex structure moduli 
and some of the K\"ahler moduli are  stabilised in these vacua. We obtain in an explicit fashion
the restrictions imposed by supersymmetry on  the flux, and the values of the fixed moduli. 
Some T-duals and Heterotic duals are also discussed, these are non-Calabi-Yau spaces.
A superpotential is constructed describing these duals.

\Date{February 2003}

\newsec{Introduction}

Compactifications of String Theory in the presence of flux have attracted 
considerable attention lately {\refs{\oldflux \DRS \TV \curio \louis \GKP 
\EvaFix \Beckerst \kst \joefrey \acfh \Beckers \GSS \Herman \Mayr  \deBoerPX 
\ferrara \dauria \lledo  \klst  \Atish \keshav \janlouis \dlust \kstt \micu 
{--} \GP}}.
These compactifications are of interest from several different points of view.
Phenomenologically they are attractive because  turning on flux typically 
leads to fewer moduli, and also because flux leads to warping which is of 
interest in the Randall Sundrum scenario, {\refs{\RS, \Herman}}.
Cosmologically they are worth studying because the resulting potential 
in moduli space could lead to interesting dynamics.  {}From a more theoretical 
view point these models enlarge the class of susy preserving vacua in 
string theory and one hopes this will  improve our  understanding 
of ${\cal N}=1$ string theory. 

In this paper, we will study type IIB string theory in the presence of flux. 
General considerations pertaining to this case were discussed in \GKP. 
Subsequently, a concrete example was studied in \kst\ involving flux 
compactifications of IIB on an $T^6/Z_2$ orientifold. In many ways this 
paper can be  thought of as an continuation  
of the investigation begun in \kst. Here we  study the  next  simplest case, 
where the compactification is on a $K3 \times T^2/Z_2$ orientifold. The 
purpose behind these  investigations is two fold. Qualitatively, one would 
like to gain some appreciation for how easy it is to obtain stable 
supersymmetric vacua after flux is turned on.  This is important, 
in view of the discussion in \GKP\ which shows that supersymmetry is 
generically broken once flux is turned on, and also bearing in mind that  
previous attempts  to turn on flux have usually  lead to unstable vacua 
with runaway behaviour \refs{\PolchinskiSM  \MichelsonPN {--} \dagata}.  
Quantitatively, one would like to know how much 
information can be obtained about the susy preserving vacua, whether 
the required non-genericity of the flux can be spelt out in an explicit 
manner, and whether  the location of the minimum in moduli space can be 
determined and the resulting masses of moduli be obtained. 

We will carry out this investigation in detail in this paper. 
We solve the supersymmetry conditions explicitly to obtain  the general 
susy preserving vacua in this model. The main  point worth emphasising 
about our analysis is that we obtain our result without having to explicitly 
parametrise the moduli space of complex and K\"ahler deformations on $K3$. 
Such a parametrisation would both be inelegant and impractical. Instead, 
by using an important theorem, called Torelli's theorem \aspin, which  
pertains to the complex structure of $K3$, we are able to map the problem 
of finding susy preserving vacua into a question of  Linear Algebra in the 
second cohomology group $H^2(K3,{\bf R})$ of $K3$. This question turns out to 
be easy to answer and yields the general susy preserving vacua mentioned above. 

Our analysis allows us to state the required conditions on the flux for a 
susy preserving vacuum in a fairly explicit manner. We find that, qualitatively 
speaking, these conditions are easy to meet, so that  several susy preserving 
vacua exist. The complex structure moduli are generically completely frozen 
in these vacua and  some but not all  of the K\"ahler moduli are also fixed. 
It is quite straightforward to determine in a quantitatively precise manner 
the location of the vacua in moduli space. 

This paper is organised as follows. \S2\  discusses some important preliminary 
material. \S3\ which contains some of the key results of the paper, describes 
the general strategy mentioned above for finding supersymmetry preserving 
vacua. This general discussion is illustrated by examples in \S4\ where 
two cases  are discussed in some length. 

The case of ${\cal N}=2$ supersymmetry  does not fall within the 
general discussion of \S3\ and is analysed  in \S5 with an example. 
The resulting moduli spaces of complex and K\"ahler deformations are also determined. 
 
In \S6 we construct an interesting infinite  family of  fluxes,
unrelated by duality,   all of which 
require the same number of D3 branes for tadpole cancellation. 
We find however that only one member of this family gives rise to an allowed vacuum. 

Various dual descriptions of the $K3 \times T^2/Z_2$ model are discussed in in \S7. One and two T-dualities
give rise to Type $I^{'}$ and Type I descriptions. The latter in turn,
after S-duality, gives rise to  models in heterotic string theory.
The resulting compactifications are non-Calabi-Yau spaces in general.  An
explicit superpotential, which is valid quite generally,  is constructed in 
these dual descriptions. It depends on
various fluxes and certain twists in the geometry.

Our methods can be used to provide a general solution for flux compactifications
 on $T^6/Z_2$ considered in \kst\ as well (building on an approach discussed in 
that paper).  This is discussed in \S8\ briefly. 

Finally, some  details   are discussed in the 
appendices A, B and C. 

Let us end by commenting that the 
  $K3 \times T^2/Z_2$ model with flux, studied in this paper,
has also  been analysed  in \keshav. Various  features of the compactification were deduced in that paper by considering an M-theory lift, 
and also the heterotic dual was discussed in some detail \dlust .   
While this manuscript was in preparation we became aware of \bbdg,
which discusses various aspects of Heterotic compactifications with flux 
in some depth.
We thank the authors for discussion and for keeping us informed of their results prior to publication.

\newsec{Background}

Some  background material for our investigation is discussed in this section. 
\S2.1\  discusses some of the essential features of IIB compactifications
with flux,
\S2.2\ gives more details relevant to the  $K3 \times T^2/Z_2$ case.    
\S2.3\ discusses some basic facts about $K3$ that will be relevant in the discussion
below. Finally \S2.4\ briefly discusses the lifting of open string moduli. 

\subsec{IIB compactifications  with Flux}

Our starting point is a compactification of IIB string theory on 
a $K3\times T^2/Z_2$ orientifold. The orientifold $Z_2$ group is given by 
$\{1,\Omega(-1)^{F_L} R \}$ where $R$ is a reflection which inverts the two coordinates of the $T^2$
and $\Omega$ and $F_L$ stand for  orientation reversal in the world sheet theory  and  fermion number in the left moving sector  respectively. 
This model is T-dual to Type I theory on 
$K3 \times T^2$ which in turn is S-dual to Heterotic on $K3 \times T^2$.

The main aim of this paper is  to study supersymmetric compactifications
after turning on various  NS-NS and R-R fluxes in this background. 
The general analysis of such flux compactifications was carried out in 
\GKP.  Let us summarise some of their main results here.

Turning on flux alters the metric of the internal space  by an overall 
warp factor. For the case at hand this means upto a conformal factor the 
internal space is still $K3\times T^2/Z_2$. 

Define 
\eqn\defg{G_{3}=F_{3}-\phi {\cal H}_{3}~,}
where $F_{3}=dC_{2}, {\cal H}_{3}=d{\cal B}_{2}$ 
denote the RR and NS three forms
and $\phi=a+ i/g_s$ denotes the axion-dilaton.   
(Our notation closely follows that of \GKP, see also \kst). 
$N=1$ supersymmetry requires that  $G_{3}$ is of type $(2,1)$
with respect to the complex structure of $K3\times T^2/Z_2$,
i.e. it has index structure $\left[G_{3}\right]_{ij{\bar k}}$ 
where $i,j$ denote holomorphic indices and ${\bar k}$ an anti 
holomorphic index. Furthermore for supersymmetry one needs 
$G_{3}$ to be primitive, that is
\eqn\defpr{J\wedge G_{3}=0}~,
where $J$ denotes the K\"ahler two-form on $K3\times T^2/Z_2~.$   

One of the main motivations for studying  compactifications with flux is moduli 
stabilisation.  The requirement that $G_{3}$ is $(2,1)$ typically completely 
fixes the complex structure for the compactification, we will see this in the 
analysis below for the example at hand. The primitivity condition \defpr\
 is automatically met if the compactification is  a (conformally) Calabi-Yau 
threefold ($CY_3$), since there are no non-trivial 5-forms on a $CY_3$.  
For $K3\times T^2$ in contrast this condition is not automatic and does 
impose some restriction on the K\"ahler moduli.  An important K\"ahler modulus
which is left unfixed is the overall volume. Since the primitivity condition 
is unchanged by an overall rescaling $J \rightarrow \lambda J$, it does not 
lift this modulus. 

One comment about non-susy vacua is also worth  making at this stage. The low-energy 
supergravity obtained after turning on flux is  of the no-scale type.  The 
flux gives rise to a potential for the moduli with minima at zero energy. The 
requirements imposed by minimising the potential are less restrictive than 
those imposed by supersymmetry.  To minimise the potential $G_{3}$ can have 
components  of $(2,1)$, $(0,3)$ and $(1,2)$ type.
The $(2,1)$ component must be primitive and the $(1,2)$ component must be of the form 
$J\wedge \alpha$ where $\alpha$ is a non-trivial $(0,1)$ form.  
We will have more to say about  non-susy vacua and  the primitivity condition
 in \S6.

\subsec{The $K3 \times T^2/Z_2$ case in more detail.}

It is worth commenting more  on some details relevant to the $K3 \times T^2/Z_2$
compactification we will study in this paper. 

The first comment   relates to tadpole cancellation for various RR fields. 
Consider first  the $7$-brane tadpole. 
The $Z_2$ orientifold symmetry has 4 fixed points on the $T^2$,
an $O7$-plane is  located at each of these
fixed points. To cancel the resulting $7-$brane charge 16 D7-branes need to be added.

Next consider  the $3$-brane tadpole \djm . Both the $O7$-planes and the 
$D7$-branes wrap the $K3$ (besides filling spacetime). As a result, it turns 
out, $2$ units of D3 brane charge are induced on the world volume of 
each $O7$-plane and one units of $D3$ brane charge in each D7-brane, 
\foot{The simplest way to see this is the following: in F-theory there 
are no $O7$ planes. Instead there are  $24$ $(p,q)$ 7-branes each of which 
acquires one unit of $3$-brane charge on wrapping $K3$. } giving rise to 
a total of $24$ units of three brane charge. This charge needs to be canceled 
by adding $D3$-branes and flux.
The relevant formula for tadpole cancellation then takes the form:
\eqn\tad{{1 \over 2} N_{flux} + N_{D3}=24 ~,}
where 
\eqn\defnflux{N_{flux}={1 \over (2\pi)^4 (\alpha^{'})^2} 
\int_{K3 \times T^2} {\cal H}_{3}\wedge F_{3}~.} 
Note in particular that with our normalisations the integral in \defnflux\ 
is over the covering manifold, before identification under the orientifold 
$Z_2$ symmetry. Actually, \tad\ is not the most general expression for the 
tadpole condition, instantons excited in the world volume of the $7$-branes 
will  give rise to $3$-brane charge in general. In this paper we will only 
deal with the case where these instantons are not excited.

The second comment pertains to  the number of moduli from the closed string 
sector present in the model before flux is turned on. In the closed string 
sector the fields $g_{MN}, C_{4},C_{0}, \phi$ are even under the action 
of $\Omega (-1)^{F_L}$, while ${\cal B}_{2},C_{2}$ are odd. This means one gets 
4 gauge bosons from ${\cal B}_{2}$ and $C_{2}$. 
In addition the metric, and $C_{4}$ 
give rise to $61$ and $25$ scalars respectively, yielding a total of 
$86$ scalars. The resulting compactification has ${\cal N}=2$ supersymmetry 
(before flux is turned on) with the gravity multiplet, three vector multiplets 
and twenty hypermultiplets. In addition, while we do not provide a precise 
count here, there are moduli that arise from the open string sector.
These include moduli due to exciting the gauge fields on the $7$ branes, 
the location of the $7$-branes on the $T^2$ and  the locations of the $D3$ 
branes in the $K3\times T^2/Z_2$. 

Finally, we discuss  the quantisation conditions which must be
met by the ${\cal H}_{3}$ and $F_{3}$ flux.
 Due to the discrete identification in the compactification this
condition is a bit subtle.
The orientifold
$K3 \times T^2/Z_2$ has additional `half' three-cycles not present in the covering
manifold $K3\times T^2$. To satisfy the quantisation condition on these three-cycles one requires
that
\eqn\quant{{1\over (2\pi)^2 \alpha^{'}}\int_\gamma F_{3} = 2 {\bf Z} ~,~
{1\over (2\pi)^2 \alpha^{'}} \int_\gamma {\cal H}_{3}= 2 {\bf Z},}
where $\gamma$ is an arbitrary class of $H_3(K3 \times T^2,{\bf Z})$.
Other possibilities which include turning on exotic flux at the $O$ planes 
were discussed in \joefrey\ but will not be explored further here.

In the discussion below we will explicitly parametrise the fluxes as follows.
Choose coordinate $x,y$ for the $T^2$, with $0\le x,y\le 1$.
The flux which is turned on must be consistent with the orientifold  $Z_2$ 
symmetry. Since ${\cal B}_{2}$ and $C_{2}$ are odd under $\Omega (-1)^{F_L}$ 
this means the allowed $F_{3}$ and ${\cal H}_{3}$ must  have two legs along the 
$K3$ and one along the $T^2$. That is 
\eqn\deffluxf{{1 \over (2 \pi)^2 \alpha^{'}} F_{3}=\alpha_x\wedge dx + \alpha_y\wedge dy}
\eqn\deffluxh{{1\over (2\pi)^2 \alpha^{'}} {\cal H}_{3}=\beta_x \wedge dx + \beta_y \wedge dy,}
where $\alpha_x,\alpha_y,\beta_x,\beta_y \in H^2(K3,{\bf Z})$.
If $e_i$ is a basis of $H^2(K3,{\bf Z})$ and $\alpha_x=\alpha_x^i e_i$ we 
have taking into account the 
quantisation condition that   $\alpha_x^i$ is 
even integer, similarly for $\alpha_y,\beta_x,\beta_y$.
Finally, we note that with our choice of orientation, as explained in App. A, 
$N_{flux}$ \defnflux,  takes the form
\eqn\tadthree{N_{flux}=\int \left(\beta_x\wedge \alpha_y\wedge dx \wedge dy 
+ \beta_y \wedge \alpha_x \wedge dy \wedge dx\right) 
= \left(-\beta_x \cdot \alpha_y + \beta_y \cdot \alpha_x\right) ~.}

\subsec{Some essential facts about  $K3$}

$K3$ is the two (complex) dimensional Calabi-Yau manifold.
It is K\"ahler and has $SU(2)$ holonomy. 
For an excellent review, see \aspin\ .

$H^2(K3,{\bf R})$ is $22$ dimensional. 
An inner product can be defined in $H^2(K3,{\bf R})$ as follows.
If  $e_i, e_j \in H^2(K3,{\bf R})$
\eqn\definner{(e_i,e_j)\equiv \int_{K3} e_i \wedge e_j.}  
This inner product matrix  can be shown to have signature $(3,19)$.
$H^2(K3,{\bf R})$ with this metric can be naturally embedded 
in ${\bf R}^{3,19}$.

$H^2(K3,{\bf Z})$, can be  thought of as a  lattice. 
It is known to be even and self dual.
These two conditions are highly restrictive.  
In a particular basis $e_i$ (discussed in App.~A) the inner product
matrix  \definner, can be shown to have the form, eq. A.3 (App.~A)
consistent with these restrictions.
We will refer to the lattice, together with this inner product, 
as $\Gamma^{3,19}$, below.

The moduli space of complex structures on $K3$ is  particularly relevant   for this paper
since many directions lifted by flux lie in this moduli space.
Torelli's theorem is  important in this context.
It says that, upto discrete identifications, the moduli space of 
complex structures on $K3$ is  given by the space of possible  
periods of the holomorphic two-form $\Omega$.

Decomposing $\Omega$ into its real and imaginary parts we have
\eqn\decome{\Omega = x+ i y}
where $x,y \in H^2(K3,{\bf R})$.
$\Omega$ satisfies two conditions
\eqn\omegaconda{\int \Omega\wedge \Omega 
               = x \cdot x - y \cdot y + 2i x \cdot y = 0 ~, }
and
\eqn\omegacondb{\int \Omega \wedge {\bar \Omega} = x \cdot x + y \cdot y >0 ~.}

{}From \omegaconda\ we see that $x,y$ are orthogonal
and from \omegacondb\ that both $x,y$ are spacelike.
So  $x,y$ span a space-like two-plane in $H^2(K3,{\bf R})$ or 
equivalently ${\bf R}^{3,19}$.
Changing the orientation of the two-plane corresponds to taking
$\Omega \leftrightarrow {\bar \Omega}$.
The space of possible periods is then the space of oriented two-planes 
in ${\bf R}^{3,19}$.
This space is called the Grassmanian,
\eqn\kthreecmod{G= O(3,19)^{+}/(O(2)\times O(1,19))^+.}
It is twenty (complex) dimensional.

Next, let us consider the K\"ahler two-form, ${\tilde J}$ on $K3$.
Since $K3$ is K\"ahler,~${\tilde J}~\in~H^2(K3,{\bf R})$.
In addition it satisfies two conditions:
\eqn\condkahlera{\int {\tilde J}\wedge \Omega =0,}
and
\eqn\condkahlerb{\int {\tilde J}\wedge {\tilde J} >0.}
With respect to the inner product, \definner, the first condition tells us 
that ${\tilde J}$ is orthogonal to $\Omega$, while the second says that it 
is a space-like two-form.

Putting this together with what we learnt above, we see that  
the choice of a  complex structure and K\"ahler two-form on $K3$   
specifies an oriented space-like three-plane $\Sigma$ in 
$H^2(K3,{\bf R})\in {\bf R}^{3,19}$. The Einstein metric is 
completely specified once this choice is made.  
It is then easy to see that  the moduli space of
 Einstein metrics on $K3$,
upto discrete identifications, is of the form
\eqn\defgrasstwo{M_E \simeq O^+(3,19)/(SO(3) \times O(19)) \times R_+.}
This is $58$ real dimensional \foot{Different choices of an oriented 
two-plane in the chosen three-plane give rise to different complex 
structures consistent with the same Einstein metric.  
This is a reflection of the underlying Hyper K\"ahler geometry.}.

Finally, we will work in the supergravity approximation in this paper.
With that in mind, it will be useful to know in the following discussion when the 
 the moduli are stabilised at a point
away from an orbifold singularity of $K3$.
Theorem 4 in \aspin\ tells us that this condition is met if
 the space-like three plane $\Sigma$ is not orthogonal to any Lattice Vector 
\foot{In terms of the basis $e_i$, a Lattice Vector is of form, $n^ie_i, n^i \in {\bf Z}$.}
 of $\Gamma^{3,19}$.

\subsec{Open String Moduli}
Most of this paper deals with closed string moduli. Before going on  though 
it is worth briefly commenting  on  open string moduli. 
These moduli are of two kinds. If D3-branes are present, their location in the 
$K3 \times T^2/Z_2$ space gives rise to moduli. Turning on flux does not freeze these
fields as long as the conditions for supersymmetry are met,  
this can be easily seen 
from the supersymmetry analysis, \granapolch. 

The second kind of open string moduli 
are  the location of the  D7- branes in the transverse $T^2$ directions. 
There are $16$ such  branes present. In the analysis below 
we  take them  to be symmetrically distributed so that 
each O7 plane has $4$ D7 branes on it. 
As a result  the dilaton will be constant 
(except for the singularities at the D7/O7 planes).
We  then consistently seek solutions  which meet the  conditions for supersymmetry,
discussed in \S2.1. 

Perturbing around such a solution, one can show that generically 
 the 7-brane moduli are lifted. 
There are two ways to see this. {}From the point of view of  ten-dimensional IIB theory,
this happens  because of the restrictive nature of the $(2,1)$ condition that  
$G_3$ must satisfy. Note that this condition must be met at every point in the 
compactification.
Once  the D7-branes move away from the O7-planes, the dilaton   varies and the $(2,1)$
 condition will not be met generically. 
{}From the F-theory viewpoint this can be understood similarly 
as a consequence of the restrictive nature of the $(2,2)$ condition that $G_4$ must satisfy. 
It is easy to see that any susy solution with constant dilaton in the IIB theory
 lifts to a susy preserving solution of F-theory with $G_4$ being of type $(2,2)$ and primitive.
Now the $(2,2)$  condition imposes strong  constraints on the complex structure moduli and 
in fact over 
determines them. As a result, perturbing around a given solution  we generically do not 
expect any complex structure moduli to 
survive. {}From the IIB perspective this means all the D7-brane moduli as well as the 
dilaton and the complex structure of the $T^2$ should be fixed. 

Let us end by commenting that some of these  issues are discussed in more detail, 
from the perspective of the Heterotic theory,
in \bbdg.

\newsec{Supersymmetric Vacua}

In the following two sections we solve the conditions imposed by 
supersymmetry and find the general supersymmetric vacua     
for flux compactifications on $K3 \times T^2/Z_2$.
Since the discussion is quite technical it is helpful to first 
summarise the key ideas.

The conditions imposed by supersymmetry are well known. 
The flux, $G_{3}$, should be of type $(2,1)$ and it
 should satisfy the requirement of primitivity.

The main challenge is to explicitly implement these conditions and find 
susy preserving vacua. The $(2,1)$ condition in particular is very 
restrictive, and, as we shall see below, cannot be met for generic fluxes. 
In finding the general supersymmetric solutions we will 
determine  both the allowed  flux and  the 
resulting susy  vacua. 
A brute force approach, relying on an explicit parametrisation of  moduli
space, is not practical for this purpose. For example, the complex structure 
moduli space of $K3$ is $20$ dimensional and not easy to explicitly  
parametrise. 

The key to making progress is Torelli's theorem for $K3$, which we discussed  
in \S2.3\ above.  This theorem allows us to restate the search for susy 
solutions as a problem in Linear Algebra in $H^2(K3,{\bf R})$. Seeking   
a complex structure in which the flux is of type $(2,1)$ translates to 
searching for an appropriate space-like two-plane in $H^2(K3,{\bf R})$.
The restrictive nature of the  $(2,1)$ condition, which we mentioned above,
can now be turned into an advantage. The flux defines a four dimensional  
subspace of $H^2(K3,{\bf R})$, which we denote as $V_{flux}$. Susy requires 
that the  two-plane must lie in $V_{flux}$. This is itself a big 
simplification since it narrows the  search from the $22$ dimensional space,  
$H^2(K3,{\bf R})$, to a four dimensional one. But in fact one can do even 
better. One finds that the $(2,1)$ condition determines the two-plane 
completely in terms of the dilaton-axion and the complex structure modulus 
of the $T^2$. The remaining conditions then determine these two moduli and 
provide some of the required  conditions on the flux.

Turning next to the requirement of primitivity, which is a condition on the 
K\"ahler two-form, one finds it  maps to the search for a space-like vector 
in $H^2(K3,{\bf R})$ which is orthogonal to $V_{flux}$.  Unlike in the case 
of the complex structure, this is not a very restrictive condition. It does 
impose some additional conditions on the flux, but once these conditions are
met, many solutions exist in which some but not all K\"ahler moduli are frozen.

In this way we determine the most general susy preserving solutions for 
flux compactifications on $K3 \times T^2/Z_2$.

The rest of this section will present the analysis sketched out above in more 
detail. In \S3.1\ we discuss  the $(2,1)$ condition. We show that it determines
 the complex structure of $K3$ in terms of the dilaton-axion and the complex 
structure of the $T^2$. We also find the constraints it imposes on these two 
moduli and on the flux. In \S3.2\ we solve these constraints explicitly and 
determine the complex structure of $K3$, $T^2$ and the dilaton-axion. We also 
make the constraints on the flux explicit. In \S3.3\ we discuss the primitivity 
condition. We determine the  restrictions it imposes on the K\"ahler moduli and 
on the flux. Finally, we summarise the results of this section in \S3.4\
stating  the resulting values of the complex structure moduli, the restrictions 
on the K\"ahler moduli, and the conditions imposed on the flux for  susy 
preserving vacua.

\subsec{The $(2,1)$ Condition}

We start by considering the restrictions imposed by the  condition 
that $G_{3}$ is of type $(2,1)$. 

One way to formulate these conditions  is to construct 
the superpotential \refs{\oldflux, \GKP}, 
\eqn\superpot{W=\int G_{3} \wedge \Omega_{3}}
which  is a function of all the complex structure moduli and also 
of the axion-dilaton. 
One can show then that  imposing the conditions:
\eqn\conda{ W=\partial_iW=0,}
 where $i$ denotes any complex structure modulus or the dilaton field,
ensures that $G_{3}$ is purely of type $(2,1)$. 
Notice  that we have one more condition than variables in \conda\ .
Two conclusions follow from this. First, as was mentioned above,
we see that  for generic fluxes there are no susy preserving minima. 
Second, we learn that in the class of fluxes which preserve susy generically 
all complex structure moduli are  stabilised. 
In finding susy  solutions we will determine below what conditions the flux 
must satisfy along with the resulting values for the complex structure
moduli and the dilaton. 

A straightforward way to proceed, as followed in \kst, is to explicitly
parametrise the complex structure moduli space, determine  $W$, and then 
search for solutions. 
For the case at hand, this is not very practical. The main complication is the 
complex structure moduli space of $K3$. An explicit parametrisation of the 
twenty dimensional Grassmanian is possible, but the resulting expressions 
are quite unwieldy. 

The crucial idea which allows us to make progress is Torelli's theorem, as was
 mentioned above. We are seeking a complex structure,   with respect to 
which $G_{3}$ is of type $(2,1)$. Torelli's theorem states that the  
complex structure is specified by a space-like two plane in $H^2(K3,{\bf R})$. 
We can think of this two-plane as determining the holomorphic two-form 
$\Omega$ and from it the complex structure. With this in mind we  
recast the search for the required complex structure in terms of conditions 
on the two-plane. Fortunately, as we see below, these conditions are simple 
to state and restrictive enough to determine the two-plane without requiring 
an explicit parametrisation of the complex structure moduli space of $K3$.

We begin by choosing the complex structure modulus on the $T^2$ to be $\tau$, 
so that we can define complex coordinates on it of the form,
\eqn\deftau{z=x+\tau y~, \  {\bar z}= x+ {\bar \tau} y ~. }
In \deffluxf,\deffluxh, we defined four vectors, 
$\{\alpha_x,\alpha_y,\beta_x,\beta_y\} 
\in H^2(K3,{\bf Z}) \subset H^2(K3,{\bf R})$  
which specify the three-form flux completely. We will refer to these  
as the flux vectors below. Together they  define  a subspace of 
$H^2(K3,{\bf R})$ which we call $V_{flux}$. $G_{3}$ can then be written as 
\eqn\defnxny{G_{3} = n_x dx + n_y dy ~,} 
where  the $2$-forms $n_x,n_y$ are given by 
\eqn\nxny{n_x \equiv \ax - \phi \bx ~, n_y \equiv \ay - \phi \by ~. }
In terms of the complex coordinates on the $T^2$, \deftau, 
we can express $G_{3}$ as 
\eqn\gthree{
G_{3} = {1\over \bar\tau - \tau}
\left(G_z dz + G_{\bar z} d{\bar z}\right) ~
}
with
\eqn\defgz{G_z = (n_x\bar\tau - n_y) }
and
\eqn\defgzbar{ G_{\bar z} = - (n_x \tau - n_y) ~.}

For $G_{3}$ to be of $(2,1)$-type the two form $G_{\bar z}$ in $K3$ must be 
of type $(2,0)$. Since the Holomorphic two-form is unique we learn that 
\eqn\cchoice{ G_{\bar z} = c\ \Omega, }
where $c$ is a constant.
We must emphasise that \cchoice\ determines the complex structure of $K3$ 
completely in terms of 
the flux, and the moduli, $\phi,\tau$ (which enter in \defgzbar) \foot{
We will assume in this section that $G_{\bar z}$ is non-zero, so that the 
constant $c$ in \cchoice is non-zero. The case where $G_{\bar z}=0$ is 
dealt with separately in  \S5.}.  
Eq.\cchoice,  will play a central role in the discussion below.

Now we can use the remaining conditions imposed by susy and consistency to determine
the remaining two moduli, the dilaton-axion and $\tau$, and also obtain the required
 conditions which should be satisfied  by the non-generic flux for a susy condition to
 exist.

{}From \omegaconda\ we see that the following condition must hold:
\eqn\conde{G_{\bar z} \cdot G_{\bar z} =0. }

Two other conditions arise as follows. 
For $G_{3}$ to be of type $(2,1)$ we see from \defgz, that 
$G_z$ must be of type $(1,1)$. This means 
\eqn\oneone{ \Omega  \cdot G_z = 0 ~,}
and
\eqn\oneoneb{ \Omega  \cdot {\bar G_z} = 0 ~, }
which can be reexpresed as 
\eqn\condc{G_{\bar z}  \cdot G_z=0}
and
\eqn\condd{G_{\bar z} \cdot {\bar G_z} = 0.}

One final condition arises from the requirement that the two-plane defining 
$\Omega$ is spacelike, \omegacondb. {}From \cchoice, it takes the form:
\eqn\condf{G_{\bar z} \cdot {\bar G_{\bar z}} >0. }

We see that, \conde,\condc, and, \condd, are three polynomial equations in
two variables, $\phi$ and $\tau$. Generically they will not have a solution.
This was expected from our discussion of the superpotential at the beginning of this section.
Two of these equations can be used to solve for 
 $\phi$, and  $\tau$. 
The third equation then   gives  two real conditions on the flux.
Finally,   we also  need to ensure that the flux meets the  inequality 
\condf.

\subsec{Solving the Equations}

In this section we  discuss in more detail how to explicitly solve the three equations,
\condd,\conde, \condf, which were obtained above from the $(2,1)$ requirement.

These equations can be written as 
\eqna\reone
$$ (\bar n_x \tau - \bar n_y) \cdot (n_x \tau - n_y) = 0 \eqno\reone a $$
$$ (n_x \bar\tau - n_y) \cdot (n_x \tau - n_y) = 0 \eqno\reone b $$
$$ (n_x \tau - n_y) \cdot (n_x \tau - n_y) = 0 ~. \eqno\reone c $$

%

We will in particular be interested in non-singular solutions for which 
${\rm Im}\phi,{\rm Im}\tau \ne 0$. 
Since, ${\rm Im}\tau \ne 0$,
\reone b, \reone c, give
\eqn\genconda{ n_x \cdot  (n_x \tau - n_y)  = 0, }
and
\eqn\gencondb{ n_y \cdot  (n_x \tau - n_y) = 0. }
Similarly, since, $Im \phi \ne 0$  \reone a, \reone c, give
\eqn\gencondc{(\ax \tau - \ay) \cdot (n_x \tau - n_y) = 0 ~.}

Using \genconda\ we can eliminate $\tau$ from the remaining two equations
to get,

\eqna\remind
$$ (n_x \cdot n_x) (n_y \cdot n_y) - (n_x \cdot n_y)^2 = 0 \eqno\remind a $$
$$ (\ax \cdot n_x) (n_x \cdot n_y)^2 
- (n_x \cdot n_x) (n_x \cdot n_y) \left( \ax \cdot n_y + \ay \cdot n_x \right) 
+ (\ay \cdot n_y) (n_x \cdot n_x)^2 = 0 ~. \eqno\remind b $$

Using the expressions for $n_x$ and $n_y$ from \nxny\ this yields two
polynomials, one quartic and the other quintic  in $\phi$, of the form:

\eqna\qurqui
$$ q_1 \phi^4 + q_2 \phi^3 + q_3 \phi^2
+ q_4 \phi + q_5 = 0~, \eqno\qurqui a $$
$$ p_1 \phi^5 + p_2 \phi^4 + p_3 \phi^3
+ p_4 \phi^2 + p_5 \phi + p_6 = 0, \eqno\qurqui b $$
where
the coefficients $q_i, p_i,$ are real and  can be expressed in terms of inner 
products of the flux vectors, as is discussed in App. B.

For a non-singular solution, $\phi$, must be complex. Since  the coefficients of the 
two polynomials are real this
means that \qurqui a, \qurqui b, must share a common quadratic factor.

In general, as is discussed in App. B this happens when the following condition
is met:
Define the matrix
\eqn\ary{
M \equiv  \left( \matrix{& p_1 & 0 & 0 & - q_1 & 0 & 0 & 0 \cr
& p_2 & p_1 & 0 & - q_2 & - q_1 & 0 & 0 \cr
& p_3 & p_2 & p_1 & - q_3 & - q_2 & - q_1 & 0 \cr
& p_4 & p_3 & p_2 & - q_4 & - q_3 & - q_2 & - q_1 \cr
& p_5 & p_4 & p_3 & - q_5 & - q_4 & - q_3 & - q_2 \cr
& p_6 & p_5 & p_4 & 0 & - q_5 & - q_4 & - q_3 \cr
& 0 & p_6 & p_5 & 0 & 0 & - q_5 & - q_4  \cr
& 0 & 0 & p_6 & 0 & 0 & 0 & -q_5
}\right) ~.}

Then \qurqui a,\qurqui b, have a common quadratic factor, if 
$M$ has a zero eigenvalue. That is if  a non-zero column vector $X$ exists  such that
\eqn\condcommon{M \cdot X \equiv M^i_j X^j=0.}
Note that since  $M$ is an  $8 \times 7$ matrix, each column of $M$ can be thought of as a 
 vector
in an $8$ dimensional vector space.
Then \condcommon, is equivalent to requiring that only $6$ of these $7$ vectors, in the 
$8$ dimensional space, are linearly independent.

Before proceeding let us emphasise that the matrix $M$  depends only  on the
 flux vectors, so the requirement of a zero eigenvalue gives rise to restrictions
on the flux which must be met for a susy solution to exist.

Once  condition \condcommon\ is met,  the resulting  common quadratic factor has the form
\eqn\commonquad{W(\phi)=  w_1 \phi^2 + w_2 \phi + w_3}
where
\eqn\valw{
 w_1 = {q_1 \over s_1} ~, ~ w_2 = {s_1 q_2 - s_2 q_1 \over s_1^2}~,
 ~ w_3 = {q_5 \over s_3} ~, }
and $s_1,s_2,s_3$ refer to the first three  components of $X$ as is explained 
in App. B.

We can now finally solve for $\phi$  by setting $W(\phi)=0$. This gives, 
\eqn\solphi{\phi={-w_2\pm{\sqrt{w_2^2-4w_1w_3}} \over 2 w_1}.}

For a nonsingular solution, the imaginary part of $\phi$,  must not vanish.
This gives the following additional conditions on the flux,
\eqn\condphip{w_2^2 < 4 w_1 w_3 ~.}

Once $\phi$ is determined, $\tau$ can be obtained from \genconda. It is given by 
\eqn\fortau{ \tau = {n_x \cdot n_y \over n_x \cdot n_x} ~. }
We should note that in case $n_x \cdot n_x=0$, \fortau, is not valid. 
Instead one can use \reone a, which yields,

\eqn\fortaub{\tau={(\bar n_x\cdot n_y + \bar n_y \cdot n_x) \pm \sqrt{
(\bar n_x\cdot n_y + \bar n_y\cdot n_x)^2 - 
4 (\bar n_x \cdot n_x) (\bar n_y \cdot n_y)} \over 2 (\bar n_x \cdot n_x)}.}
(Requiring $Im(\tau)>0$ fixes the sign ambiguity in \fortaub.)
Eq. \fortaub, is also useful for stating an additional  condition on the 
flux which arises from the requirement that $Im\tau \ne 0$. This  condition
takes the form

\eqn\condtaup{({\bar n}_x \cdot n_y 
+ {\bar n}_y \cdot n_x)^2 < 4 ({\bar n}_x \cdot n_x)
({\bar n}_y \cdot n_y) ~,}
where $\phi$, is given in \solphi.
 
Finally a condition on the flux arises from \condf. 
Using \cchoice, this takes the form, 
\eqn\positive{ ( n_x \tau - n_y ) \cdot ({\bar n_x} {\bar \tau} -{\bar n_y}) > 0 ~, }
with $\tau$ and $\phi$ given in terms of the flux, in \fortau, \solphi. 

In summary,  the $(2,1)$ condition determines the complex structure of the $K3 \times T^2/Z_2$
space completely in a susy preserving solution. 
The dilaton-axion, $\phi$, and the complex structure modulus of the $T^2$, $\tau$,
 are  given by \solphi, \fortau.
The complex structure of $K3$ is determined
implicitly by $\Omega$ which is given by \cchoice.
In addition the following conditions are imposed on the flux:
the matrix $M$ \ary\ must have a zero eigenvector, and the conditions \condphip, \condtaup,
and \positive, must hold.

\subsec{Primitivity}
We turn next to the requirements imposed by the primitivity condition. 

On $K3\times T^2$, the K\"ahler two-form, $J$, is given by \foot{For 
ease in subsequent discussion, we drop the overall factor of $i$,  
that is  conventionally present in the definition of $J$.}  
\eqn\formj{J={\tilde J} + g_{z{\bar z}}dz\wedge {\bar dz},}
where $z,{\bar z}$ denote coordinates on the $T^2$ and 
${\tilde J} \in H^{(1,1)}(K3)$. 

The primitivity condition is \defpr.
{}From the form of $G_{3}$ we see that there are no constraints 
on $g_{z{\bar z}}$, so the K\"ahler modulus of the $T^2$ is not fixed. 
The constraints on ${\tilde J}$, 
in terms of the inner product, \definner,  take the form,
\eqn\condprima{{\tilde J}\cdot G_z = {\tilde J} \cdot G_{\bar z} =0.}
Since $Im(\tau), Im(\phi)\ne 0,$ for a non-singular solution, we learn from 
\nxny\ that   
\eqn\primc{{\tilde J} \cdot \alpha_x={\tilde J} \cdot \alpha_y=
{\tilde J} \cdot \beta_x= {\tilde J} 
\cdot \beta_y=0.}
Thus  ${\tilde J}$ must be orthogonal to the vector space $V_{flux}$.

An acceptable ${\tilde J}$ must meet the following two  additional conditions. 
It should be a space-like, i.e., positive norm vector in $H^2(K3,{\bf R})$. 
And, it should be of type $(1,1)$. This latter condition can be stated as 
follows: 
\eqn\oneonej{{\tilde  J} \cdot \Omega = {\tilde J} \cdot {\bar \Omega}=0.}
{}From \cchoice, we see that \oneonej, is automatically met if \primc, is true. 
This leaves the two conditions of ${\tilde J}$ being space-like and 
orthogonal to $V_{flux}$. Since $H^2(K3,{\bf R})$ is $22$ dimensional, 
at first glance, it would seem that these  conditions can be met for 
generic fluxes, leaving $18$ of the $K3$ K\"ahler moduli  unfixed.

Some thought shows that this is not true and that in fact the  flux must meet 
some conditions in general. The metric on $H^2(K3,{\bf R})$ gives rise to an 
inner product matrix in $V_{flux}$ in an obvious manner.  
It is convenient to state the 
restrictions on the flux, in terms of the number of non-trivial eigenvectors, with positive, 
negative and null norm, of this matrix. 
One can show, as we will argue below, that the 
number of positive norm  eigenvectors must be $2$ and the number of null eigenvectors must be 
$0$, for  a non-trivial space-like ${\tilde J}$ to exist that meets, \primc.
This leaves the following three possibilities:

a) dim $V_{flux} = 2$, $(2 +,0 -)$

b) dim $V_{flux}= 3$, $(2 +,1 -)$

c) dim $V_{flux}= 4$,  $(2 +,2 -),$

\noindent
where in our notation $(2 +, 1 -)$ means two  eigenvectors of positive norm
and one of negative norm etc.

In the next two paragraphs we pause to discuss  how this conclusion comes 
about. Thereafter we return to the main thread of the argument again. 
Since $\Omega$ lies in $V_{flux}$ and is space-like, the number of 
positive-norm eigenvectors must be at least two. Since the signature 
of $H^2(K3,{\bf R})$ is $(3,19)$ the maximum number of positive norm 
eigenvectors can be three. But if it is three then ${\tilde J}$ cannot 
be orthogonal to $V_{flux}$ and still be spacelike \foot{One can construct 
a basis of $H^2(K3,{\bf R})$, starting from these three spacelike vectors 
and appending $19$ time-like vectors to them. Then if ${\tilde J}$ is 
orthogonal to the three spacelike ones it must be purely time like.}. 
Thus, there must be exactly two  spacelike eigenvectors in $V_{flux}$. 

Next, we  turn to the number of eigenvectors with null norm.
Any such eigenvector must be orthogonal to the two eigenvectors with 
positive eigenvalues. So if such an eigenvector exists, and ${\tilde J}$ 
is orthogonal to it, besides being orthogonal to the two space-like vectors, 
one can again show that it cannot be space-like \foot{Let the two spacelike 
vectors be $v_1,v_2$ and the null vector be $ v_N=v_3+v_4,$ where 
$v_3$ is spacelike and $v_4$ is time like. Then orthogonality would require 
that ${\tilde J}= v_N + v_t$, where $v_t$ is a time like vector orthogonal 
to $v_1,v_2,v_N$. This makes ${\tilde J}$ time-like.}.

The argument in the  previous two paragraphs shows that $V_{flux}$ must meet one of the three
possibilities discussed above.  
Once the flux meets this requirement, the condition \primc, can be satisfied by a space-like
 ${\tilde J}$. Since \primc,  imposes $4$ conditions this leaves $18$ moduli in the $K3$ K\"ahler 
moduli space, besides the K\"ahler modulus of the $T^2$, unfixed. 

{\it Orbifold Singularities}

There is one final point we should discuss in this section.
This is concerned with the  existence of an orbifold singularity on the $K3$ surface.
At such a singularity various  two-cycles shrink to zero size
 and additional light states obtained by branes wrapping such cycles can enter the low energy 
theory. Since we work with a supergravity theory without these states, our analysis of the 
resulting vacuum is self-consistent only if it does not contain any such light state.

Usually in string theory at an orbifold point  additional light states are avoided by
 giving an expectation value
to the axionic  partner of the blow-up mode of  the  vanishing two-cycle. 
In our constructions we do not always have the freedom to turn on such an expectation value,
since the K\"ahler mode corresponding to blowing up the  cycle is sometimes lifted.
In such cases the axionic partner is also lifted, typically this happens because it is eaten 
by some Gauge Boson, in the low-energy gauged supergravity. 
It is an interesting question to ask what are the masses of states which arise from 
wrapped branes in such a  case, but we have not investigated this yet. 

To avoid these complications we will mainly only consider vacua below  which lie away from an 
orbifold singularity or if at an orbifold, where the relevant blow up modes are not lifted. 
Two exceptions to this are  the   discussion towards the end of  \S4.1\ which describes 
a solution generating technique that could have wider applicability, 
and  \S5\ which discusses ${\cal N}=2$ solutions. The ${\cal N}=2$  vacua could prove useful
in determining modifications to the BPS formulae for wrapped branes in the  gauged supergravity 
obtained after turning on flux. 

The rest of this subsection analyses how to determine if a solution contains an orbifold 
singularity. 

One can show that the susy preserving conditions  discussed above result in an orbifold
singularity iff
$V_{flux}$ contains a Lattice Vector $v$ of  $\Gamma^{3,19}$,  
which is orthogonal to $\Omega$.
If no such Lattice Vector exists, one can always choose the K\"ahler two form
${\tilde J}$ consistent with the primitivity conditions, to avoid an orbifold.

To see this we recall from \S2\ that the $K3$ surface is at an orbifold 
point in its moduli space if a Lattice Vector of $\Gamma^{3,19}$ exists
which is orthogonal to the three plane $\Sigma$, that  determines the 
Einstein Metric. Now if a Lattice Vector $v$ exists which is orthogonal 
to $\Omega$ and  $v \in V_{flux}$, it must be orthogonal to ${\tilde J}$ 
(due to the primitivity condition \primc\ ). Thus it must be orthogonal 
to $\Sigma$, so as claimed above the $K3$ surface is at an orbifold 
singularity. For the converse we need to consider two possibilities. 
Either there exists no Lattice Vector orthogonal to $\Omega$, in this 
case we are done. Or there is such a Lattice Vector $v$ but it does not 
lie in $V_{flux}$. In this case one can always arrange that  ${\tilde J}$,  
consistent with the condition, \primc, is not orthogonal to $v$, so again 
an orbifold is avoided \foot{Decompose $v= v_{\parallel}+ v_{\perp}$, 
where $v_{\parallel}$ lies in $V_{flux}$, and  $v_{\perp}$ is perpendicular 
to $V_{flux}$. By orienting ${\tilde J}$ to have a ("small enough") component 
along  $v_{\perp}$ one can then  ensure that ${\tilde J} \cdot v \ne 0$, 
while \primc, and positivity of ${\tilde J}$ are  met.}.

\subsec{Summary of Conditions Leading to Susy Solutions}
The discussion of this section has been quite technical and detailed. It is 
therefore useful to summarise the main results here  for further reference. 

Susy is generically broken for flux compactifications of IIB theory on 
$K3 \times T^2/Z_2$. To preserve susy the flux must meet the following 
conditions:
First, the vector space $V_{flux}$ spanned by the flux must be of one 
of the three classes, a,b,or c, discussed above in the primitivity section. 
Second, the matrix $M$, \ary, formed from the flux,  
must have a zero eigenvalue.

Once these conditions are met a solution exists. The dilaton, $\phi$, 
is given by \solphi, and in terms of $\phi$ the complex structure modulus 
of the $T^2$, $\tau$, is given by \fortau. The complex structure of the 
$K3$ is implicitly determined by $\Omega$ which is give by \cchoice\ in 
terms of $\phi,\tau$. Unlike the complex structure, the K\"ahler moduli 
are not completely determined. Instead the K\"ahler two-form of $K3$ must 
meet four conditions, \primc. This generically leaves an $18$ dimensional 
subspace of the $K3$ K\"ahler moduli space, and the K\"ahler modulus of the 
$T^2$, unfixed. 

Finally, in order to ensure that the resulting solution is non-singular 
some additional conditions must be met by the flux. To ensure that the  
complex structure moduli are stabilised at non-singular values, the 
inequalities, \positive, \condphip, and \condtaup,  must be met. And to ensure 
that orbifold singularities are avoided in the resulting solution, 
$V_{flux}$ must not contain any Lattice Vector of  
$\Gamma^{3,19},$ which is   orthogonal to  $\Omega$.

One final comment. Our approach to finding susy preserving solutions above 
relied crucially on the  fact that the constant $c$ in \cchoice, did not 
vanish. If this constant is zero the holomorphic two-form of $K3$ is not 
constrained to lie in $V_{flux}$.  This second branch of solutions will 
be considered further in  \S5,
it gives rise to ${\cal N}=2$ susy preserving vacua.

\newsec{Some Examples}

Here we will illustrate the general discussion of the preceding section 
with a few examples. \S4.1\ discusses the case where $V_{flux}$ is two 
dimensional. We consider the general solution, some examples and also 
discuss a method of generating additional susy preserving solutions 
starting from an existing one. \S4.2\ applies the general discussion 
above to an  example where $V_{flux}$ is four dimensional.

\subsec{$V_{flux}$ has Dimension $2$:}

In this case  only two of the four flux vectors are linearly independent. 
Since $\Omega$ must be a two-plane contained in $V_{flux}$, this means both 
vectors spanning $V_{flux}$ must be spacelike, so that $ V_{flux}$ is of type
$(2+,0-)$. In this case the two-plane defined by $\Omega$ must be $V_{flux}$. 

Going back to the conditions for $G_{3}$ to be of type $(2,1)$ one finds 
from \oneone, and \oneoneb, that 
\eqn\twoa{G_z \cdot \Omega = G_z \cdot {\bar\Omega} =0~. }
Since, in this case, $V_{flux}$ is spanned by $\Omega,{\bar \Omega}$, 
this means from \twoa,  that 
\eqn\dimtca{G_z=n_x{\bar \tau} -n_y=0.}
Since $Im \tau \ne 0$ for a non-singular solution, we learn from  \conde, that 
\eqn\dintcb{n_x\cdot n_x=0,}
which can be rewritten as
$$ \axx - 2 \gxx \phi + \bxx \phi^2 = 0 ~. $$
Solving for $\phi$ we obtain
\eqn\valphi{\phi = {1\over \bxx} \left( \gxx \pm \sqrt{\gxx^2 -  \axx \bxx}
\right)} 
$\Im\phi \ne 0 $ implies $\gxx^2 < \axx \bxx ~. $ So the
vectors $\ax$ and $\bx$ must be linearly independent.

With $\phi$ fixed, \dimtca, is two complex equations   in $\tau$, one of these 
can be used to determine $\tau$, the other then gives two real conditions on 
the flux. Multiplying both sides of \dimtca\ by $\bar n_x$ we have that 
\eqn\valtaua{{\bar \tau}= {n_y \cdot {\bar n_x} \over n_x\cdot {\bar n_x}} ~,}
or equivalently
\eqn\valtaub{\tau= {{\bar n_y} \cdot n_x \over n_x\cdot {\bar n_x}} ~.}
Multiplying \dimtca\ by $n_x$ gives 
\eqn\valtauc{n_x \cdot n_y =0 ~,}
substituting for $\phi,\tau$ from  \valphi, \valtaub, in \valtauc, we get the 
two conditions on the flux mentioned above. Finally we note that since 
$\Omega$, is a spacelike two plane by construction in this case, 
the inequality \positive\ is automatically met. Also since $\Omega$ spans
$V_{flux}$, the moduli of $K3$ can be chosen to be away from an orbifold point.  

Next we turn  to primitivity \primc, which requires that $\tilde J$ is 
orthogonal to the vector space $V_{flux}$. In the present example, this 
condition does not yield any extra restrictions on the flux. It can be met 
easily. $H^2(K3,{\bf R})$ is $22$ dimensional. \primc, imposes two conditions 
allowing for all twenty K\"ahler moduli of $K3$,
and the one K\"ahler modulus of the $T^2$, to vary.

As a concrete example consider the  case where $\alpha_x, \beta_x $ are the 
two linearly independent flux vectors, with 
\eqn\fluxs{\alpha_y=-\beta_x  \ \  {\rm and}  \ \beta_y=\alpha_x ~.}
In addition take the flux to satisfy the conditions 
\eqn\cfluxa{\alpha_x^2=\beta_x^2 ~,}
and
\eqn\cfluxb{\alpha_x\cdot \beta_x = 0 ~.}

{}From \valphi, one finds then that the dilaton is stabilised at the value
\eqn\valphie{\phi=\pm i ~.}
Taking $\phi=i$ from \valtaub, we have that 
\eqn\valphif{\tau=i ~.}
We see that with this choice of flux, \valtauc, is automatically met. 
As mentioned above in this case, $\Omega$ corresponds to the two plane 
spanned by $\alpha_x,\beta_x$. 

Next  we come to the tadpole conditions. 
{}From \tadthree, and \fluxs,  we see that $N_{flux}$ is given by 
\eqn\valfluxc{N_{flux}=2 \alpha_x^2,}
so that the D3 brane tadpole condition takes the form
\eqn\valtadpole{\alpha_x^2 + N_{D3} = 24 ~.}
This condition can be easily met. 

As a specific illustration, we take,  in the notation of App. A,  
\eqn\exc{\alpha_x =  2 e_1 \ , \  \beta_x =  2 e_2 \ , }
 so that conditions 
\cfluxa,\cfluxb,   are met
(the even coefficients in \exc, ensure the correct quantisation conditions).
We then have that 
\eqn\tgg{N_{D3}=24-8=16,}
number of branes must be added in the compactification. 
Let us end by mentioning  that in this example, \exc,
primitivity requires the K\"ahler two-form of $K3$ to be of the form
$$ {\tilde J} =  \sum_{i=3}^{22} t_i e_i \ , $$
where the real parameters $t_i$ are chosen to make $J \cdot J > 0 $.
This is a twenty dimensional space, as was mentioned above.

{\it New Solutions from Old}

The case where ${\rm dim}(V_{flux})=2$ also allows us to illustrate a trick 
which is sometimes helpful in finding additional solutions to the susy 
conditions. The idea is as follows:
Given a set of flux vectors which give rise to a susy solution, one can try 
to alter the flux vectors in such a manner that we  keep the dilaton and 
complex structure of both the $K3$ and $T^2$ unchanged. In particular this 
means keeping $G_{\bar z}$ unchanged \cchoice.  
Let the $\alpha_x \rightarrow \alpha_x+\delta \alpha_x$ etc. 
Then we have that 
\eqn\chnga{\delta G_{\bar z}=\delta n_x \tau -\delta n_y =0.}
Since $G_z$ must still be of type $(1,1)$ for preserving susy we have that 
\eqn\chngb{\delta G_{z} \cdot \Omega = \delta G_{z} \cdot {\bar \Omega}=0.}
This yields from \cchoice, \chnga, that 
\eqn\chngc{\delta n_x \cdot \Omega = \delta n_x \cdot {\bar \Omega}=0.}
If a $\delta n_x$ can be found, consistent with the quantisation conditions on the flux, \quant, 
that satisfies \chngc, then \chnga can be solved for $\delta n_y$.
In some cases, as we now illustrate, $\delta n_y$ is also consistent with the quantisation 
conditions. In this case, subject to the primitivity condition and the 
$D3$ brane tadpole condition \tad\ also being met, one can obtain additional susy preserving 
solutions. 

As an example consider the set of flux, \fluxs, discussed in the previous section.
In this example to begin with, $dim. V_{flux}=2$, and $\phi=\tau=i$ in the susy vacuum.
 Now suppose the flux is changed so that 
$\delta \alpha_x, \delta \beta_x$ are both orthogonal to $\alpha_x,\beta_x$.
In addition we assume that  
\eqn\chngd{\delta \alpha_y=\delta \beta_x,}
and
\eqn\chnge{\delta \beta_y=-\delta \alpha_x.}
It is easy to then see that both \chnga\ and  \chngb\ are met. 
For the new flux ${\rm dim} V_{flux} >2$, so the primitivity condition can also  be met if
$\delta \alpha_x,\delta \beta_x$ are both time like. 

Let us end with two comments.

First, the change in $N_{flux}$ is given by 
\eqn\chngf{\delta N_{flux}=-(\delta \beta_x \cdot \delta \alpha_y) + \delta \beta_y \cdot 
\delta \alpha_x}
i.e.,
\eqn\chngg{\delta N_{flux}=-\delta \beta_x^2 - \delta \alpha_x^2.}
For time like $\delta \beta_x, \delta \alpha_x $ this is positive. 
 So by altering the flux in this manner the 
number of D3 branes that need to be added can be reduced, \tad. 
In particular one can easily find  choices of $\delta \alpha_x, \delta \beta_x$  which 
give rise to a vacuum where $N_{flux}=24$ and no D3-branes need be added.

Second,  the above examples with ${\rm dim}V_{flux} >2$, 
which are generated from the old solutions by altering the flux vectors, 
correspond to orbifold singularities. This follows from the discussions on 
orbifold singularities in \S3.3 and because of the fact that the lattice 
vectors $\delta\alpha_x, \delta\alpha_y\in V_{flux}$ are orthogonal 
to $\Omega$.

\subsec{A solution with common quadratic}

\def\axxn{\alpha_{xx}}
\def\axyn{\alpha_{xy}}
\def\ayyn{\alpha_{yy}}
\def\bxxn{\beta_{xx}}
\def\bxyn{\beta_{xy}}
\def\byyn{\beta_{yy}}
\def\gxxn{\gamma_{xx}}
\def\gyyn{\gamma_{yy}}
\def\gxyn{\gamma_{xy}}

We now find some solutions of the quartic \qurqui a\ and quintic \qurqui b\ . 
For simplicity we restrict the flux vectors to satisfy
\eqn\flxansz{
\axy = \bxy = (\gxy + \gyx) = 0 \ .
}
We further assume
\eqn\flxansza{\eqalign{
& \bxx = 2 \axx = 2 \gxx \cr
& \byy = 2 \ayy = 2 \gyy ~.
}}
With this assumption the quartic and quintic reduces to
\eqn\simplqur{\eqalign{
& \axxn \ayyn \left( 4 \phi^4 - 8 \phi^3 
              + 8 \phi^2 - 4 \phi + 1 \right) = 0 ~, \cr
& \axxn^2 \ayyn \left( - 4 \phi^5 + 12 \phi^4 - 16 \phi^3 
             + 12 \phi^2 - 5 \phi + 1 \right) = 0 ~, 
}}
where $\axxn$ and $\ayyn$ are defined in App. B.
The above polynomials can be rewritten as 
\eqn\rwrtpoly{\eqalign{
& \axxn \ayyn \left( 2 \phi^2 - 2 \phi + 1 \right)^2 = 0 ~, \cr
& \axxn^2 \ayyn (1 - \phi) \left( 2 \phi^2 - 2 \phi + 1 \right)^2 = 0 ~.
}}
Clearly, they share a common quadratic
\eqn\cmnq{
2 \phi^2 - 2 \phi + 1 
}
which can be set to zero to obtain solution for $\phi$ as
\eqn\slnphi{
\phi = {1\over 2} ( 1 \pm i) ~.
}
{}From \fortaub\ and on using the assumptions \flxansz,\flxansza\ the 
expression for $\tau$ becomes
\eqn\tauvlu{
\tau = i \sqrt{\ayyn\over\axxn} ~.
}
The conditions \flxansz,\flxansza\ can be met by some suitable choice of the
flux vectors. For example consider
\eqn\fvchoice{\eqalign{
& \alpha_x = 2 (e_1 - e_2) ~, \cr
& \alpha_y = 2 (e_1 + e_2 + e_4) ~, \cr
& \beta_x = - 4 e_2 ~, \cr
& \beta_y = 2 (2 e_1 + e_4 + e_5) ~.
}}
This is a solution of $(2+,2-)$-type. For these flux vectors $\axxn = 16 $
and $\ayyn = 8$. Hence, we have $\tau = i/\sqrt{2}~.$ 
The resulting solution is non-singular. $Im(\phi),Im(\tau) \ne 0$,
and one can show that \positive, is also met.
One can also show that orbifold singularities are 
avoided. Since $Im(\tau)$ is irrational, there is no element of $\Gamma^{3,19}$ contained 
in $V_{flux}$ which is orthogonal to $\Omega$.

Finally, we note that  the contribution due to the flux  to the D3-brane 
tadpole condition, \tadthree, 
is given by
\eqn\tpoltt{
N_{flux}=\alpha_x.\beta_y - \beta_x.\alpha_y = 32 ~.
}
As a result $8$ $D3$ branes need to be added for a consistent solution.

\newsec{The Second Branch and ${\cal N}=2$ Supersymmetry}

As was mentioned at the very end of \S3\ the general strategy discussed therein  
for finding susy solutions is applicable only if the constant $c~,$ \cchoice, 
is non-zero. In  section \S5.1\ we discuss how to find solutions for which 
this constant vanishes, by  formulating both the conditions on the flux for 
such solutions to exist and  
determining the constraints on the moduli in the resulting vacua. 
We will refer to these solutions as lying in the ``second branch". In \S5.2\ we show that 
the second branch in fact   meets the necessary and sufficient conditions for preserving
${\cal N}=2$ supersymmetry. In \S5.3\ we give  an example of such a solution. 

\subsec{The Second Branch}

We begin by noting that if $c$, the constant  in \cchoice, vanishes, then 
\eqn\czero{ G_{\bar z} = n_x \tau - n_y = 0 ~. }
Equating the real and imaginary parts of \czero\
separately to zero we obtain
$$\ax \Re(\tau) - \bx \Re(\tau\phi) = \ay - \by \Re(\phi) \ , $$
$$\ax \Im(\tau) - \bx \Im(\tau\phi) =  - \by \Im(\phi) \ . $$
Thus only two of the flux vectors $\alpha_x,\alpha_y,\beta_x,\beta_y$, at best, are 
 linearly independent.
So the first thing we learn is that for a solution of this kind,
dim $V_{flux} \le 2$.
Since in  a nonsingular solution, $Im(\phi)$, does not vanish we can take these two 
independent flux vectors to be $\alpha_x,$ and $\beta_x$. 

Next let us consider the constraints coming from primitivity.
For the solutions of \S3\ this was discussed in \S3.3\ and much of that analysis   goes over to the 
present case as well. In particular, one finds again that  
${\tilde J}$ must be spacelike and orthogonal to $V_{flux}$. 

The remaining constraints come from the $(2,1)$ condition. 
This takes the form of the following equations:   
\eqna\fsttwo
 $$  G_z \cdot \Omega = 0~, ~~ G_z \cdot \bar\Omega = 0~, \eqno\fsttwo a $$
 $$  \Omega \cdot \Omega = 0~, \  
     \Omega \cdot \bar\Omega > 0 ~. \eqno\fsttwo b $$

{}From \fsttwo{a}\ get
$$ (n_x \bar\tau - n_y) \cdot \Omega = 0~, 
~~ (n_x \bar\tau - n_y) \cdot {\bar \Omega} = 0~. $$
 Using the fact, \czero\  that $n_y = \tau n_x$, and $Im(\tau) \ne 0$,
 the above conditions
can be written as
$$n_x \cdot \Omega = 0 ~, ~~ \bar n_x \cdot \Omega = 0$$
or equivalently
\eqn\sba{\ax \cdot \Omega = 0 ~, ~~ \bx \cdot \Omega = 0 ~.} 
As a result we see that $\Omega$ must be orthogonal to  $V_{flux}$. 

In \S3.3\ we found it useful to classify $V_{flux}$ by the number of positive, negative, and null norm 
eigenvectors of the inner product matrix. In the present case, putting the constraints from the primitivity and the $(2,1)$ conditions together one can show that $V_{flux}$ cannot contain any 
eigenvectors of  positive, or null norm. 
 
The argument is as follows. 
We saw in \S2.3\ that $\Omega, {\tilde J}$
together define a space-like three-plane, $\Sigma$,  in $H^2(K3,R)$. 
Now let $v_1 \in V_{flux}$ be a positive norm eigenvector,
then  $\Omega$ and ${\tilde J}$ and therefore $\Sigma$, 
 must  be orthogonal to it.
But since $H^2(K3,R)$ has  signature $(3,19)$, such a 
three-plane cannot exist.  Thus a non-singular $\Omega, {\tilde J},$ requires
that $V_{flux}$ contains no positive norm eigenvector. 
A similar argument shows that $V_{flux}$ cannot contain any null norm 
eigenvector either \foot{If $v_N \in V_{flux}$ is a null eigenvector,
we can write $v_N=v_1+ v_4$, where $v_1\cdot v_1 >0, v_4 \cdot v_4 <0, v_1 \cdot v_4 =0.$
A basis of orthogonal vectors in  $H^{(K3,R)}$ can be now constructed, 
${\cal B}=\{v_1,v_4,v_2,v_3,v_5 \cdots v_{22} \}$,
where $v_1,v_2,v_3$ are spacelike and the rest are timelike. 
Define ${\hat V}$ as the subspace spanned by the basis elements $\{v_2,v_3,v_5, \cdots v_{22} \}$.
One can show that the existence of non-singular $\Omega, {\tilde J},$ requires
 the existence of a spacelike three-plane in ${\hat V}$.
 This is impossible since ${\hat V}$ has signature $(2,18)$. }.

The only possibilities we are then left with is that $V_{flux}$ is of dim. $2$ and  type $(0+,2-)$, or of
dim $1$ and type $(0+,1-)$.  Once the flux meets these conditions susy preserving vacua can be found. 
The complex structure and K\"ahler two-form are somewhat constrained in these vacua but not completely 
fixed.
$\Omega$ is  orthogonal to $V_{flux}$ and is defined by  an oriented spacelike two plane
in the subspace $V_{\perp}$ orthogonal to $V_{flux}$, while  $\Omega, {\tilde J}$  together 
define a space-like three 
plane in $V_{\perp}$. 
E.g., in the case where dim. $V_{flux}=2$, the space of complex structures of K3 is given (upto 
discrete identifications) by the Grassmanian
\eqn\sbgrassa{G=O^+(3,17)/ (O(2) \times O(1,17))^+,}
which is $36$ dimensional, while the moduli space of Einstein metrics  has the form
(again upto discrete identifications),
\eqn\sbgrassb{M_E = O^+(3,17)/(SO(3) \times O(17)) \times R_+,}
which is $52$ dimensional. 

Also, while the K\"ahler modulus of the $T^2$ is not constrained.
the dilaton $\phi$ and the complex structure of the $T^2$, $\tau$, can be determined from \czero. 
For example, by taking the projections of \czero, along $\alpha_x$, we find $\tau$ is given by 
\eqn\sbtau{\tau={(\alpha_x \cdot n_y) \over (\alpha_x \cdot n_x)}.}
Taking a projection of \czero\  along $\beta_x$ and using \sbtau, we then get,
\eqn\sbphi{(\beta_x \cdot n_x) (\alpha_x \cdot n_y) - (\beta_x \cdot n_y)(\alpha_x \cdot n_x)=0,}
which is a quadratic equation in $\phi$ that can be solved. 

To summarise, for a solution along this second branch to exist, the following conditions must be met
by the flux:
dim. $V_{flux} \le 2$, and $V_{flux}$ must be spanned by 
 time like  vectors \foot{There are some additional constraints on the flux which come from requiring that $Im(\phi)$ and $Im(\tau)$ are non zero, these can be deduced in a straightforward manner and we will not determine them explicitly here.}. 
The dilaton and $\tau$ are then determined by \sbphi, \sbtau.  
The complex structure  and K\"ahler moduli of $K3$ are somewhat constrained but not determined completely,
and the K\"ahler modulus of the $T^2$ is not constrained at all.
    
Before proceeding further, we would like to mention that all the solutions 
in the second branch correspond to orbifold singularities. {}From \S2.3 we
learn that there exists an orbifold singularity when ever the space like 
three plane $\Sigma$ is orthogonal to a lattice vector of $\Gamma^{3,19}$. 
Primitivity requires the flux vectors to be orthogonal to the K\"ahler form 
${\tilde J}$. {}From \sba\ we find that they are also orthogonal to $\Omega$. 
Hence the three plane $\Sigma$ is orthogonal to $V_{flux}$, resulting in  
orbifold singularities.
  
\subsec{${\cal N}=2$ Supersymmetry}
We will now show that solutions in the second branch discussed above
 meet the necessary and sufficient conditions
for ${\cal N}=2$ supersymmetry. 

These conditions were discussed for the $T^6/Z_2$ case in \kst. 
A similar  analysis can be carried out  for  $K3 \times T^2/Z_2$ as well.
Here we will skip some of the details and state the main results. 
The necessary and sufficient conditions for ${\cal N}=2$ susy are the  following: 
An $SO(4) \times U(1)$ group of rotations acts on the tangent space of $K3 \times T^2$. 
For preserving ${\cal N}=2$ susy $G_{3}$ must transform as a $(3,0)_{+2}$ representation under
$SU(2)_L \times SU(2)_R \times U(1) \simeq SO(4) \times U(1)$.
This means in the notation of this paper that $G_{\bar z}$ must vanish
and $G_{z}$ must transform as the anti self-dual representation of $SO(4)$ \foot{
In \kst\ this representation was referred to as the self-dual representation. 
The discrepancy is due to an opposite choice of orientation, or equivalently opposite
choice of sign for $\epsilon_{abcd}$ (notation of \kst), in the two papers. The choice in this
paper agrees with the conventional one, \aspin\  for $K3$.}.

Since $G_{\bar z}$ must vanish we see that all ${\cal N}=2$ preserving solutions must 
lie in the second branch. We now show that all solutions in the second branch also
meet the requirement of $G_z$ being anti-self dual. 
To see this, we have to simply note that any vector belonging to $H^2(K3,R)$
 which is orthogonal to both 
$\Omega$ and ${\tilde J}$ must be an  anti-self dual two-form \foot{$H^2(K3,R)$ can be decomposed into, 
$H^{+} + H^{-}$,  the self-dual and anti self-dual subspace. $\Sigma$, the three plane formed by 
$\Omega, {\tilde J},$ is identical to $H^+$. Thus any vector orthogonal to $\Omega, {\tilde J},$ 
must be anti self-dual.}. We saw above that 
$G_z$ meets this condition  in the second branch.  
This proves that all solutions in the second branch meet the necessary and sufficient conditions
for ${\cal N}=2$ supersymmetry. 

A final comment. One should be able to associate more than one complex structure,
which still keeps $G_3$ of type $(2,1)$, with a solution of ${\cal N}=2$ susy. 
For solutions in the second branch such an additional complex structure is given by 
taking $\Omega \leftrightarrow {\bar \Omega}$. This clearly changes the  complex structure.
And from the discussion of the second branch  above it is easy to see that $G_3$ still 
continues to meet the $(2,1)$ condition \foot{$G_z$ is orthogonal to ${\bar \Omega}$ and 
$G_{\bar z}$  vanishes.} (primitivity is of course still true, since ${\tilde J}$ is unchanged).

\subsec{An Example}
For an example we consider the case where 
\eqn\exnta{\eqalign{
{1\over (2\pi)^2\alpha'} F_3=& 2 e_4 \wedge dy \ , \cr
{1\over (2\pi)^2\alpha'} {\cal H}_3=& 2 e_4 \wedge dx \ .}}
where  $e_4 \in \Gamma^{3,3} \subset \Gamma^{3,19}$ (see App.~A).
And $e_4 \cdot e_4=-2$, so that $e_4$ is a time-like vector.

{}From, \deffluxf, \deffluxh, we see that $\alpha_x=0, \beta_x=2 e_4, \alpha_y=2 e_4, \beta_y=0$.
So  $V_{flux}$  is one dimensional and is of type $(0,1-)$, i.e., it is spanned by a time like vector. 
Thus the required conditions for a solution in the second branch are met. 
It is easy to see that in this simple case, $\phi,\tau$ are not completely fixed. Rather, \czero\ imposes
 one condition on them
\eqn\exntb{ \phi \tau=-1.}
 
The moduli space of complex structure of $K3$, is now the Grassmanian 
$O(3,18)/(O(2) \times O(1,18))$
(upto discrete identifications), which is $38$ dimensional and the moduli space of Einstein metrics
on $K3$ has the form $O(3,18)/(O(3) \times O(18)) \times R_{+}$,
( again upto discrete identifications), which is $55$
 dimensional. 

The flux contribution to three brane charge is $N_{flux}/2 = 4$, 
\tadthree, \tad, so that $20$ D3-branes need to be added in this case. 
It follows from the discussion in the previous subsection that this model has 
${\cal N}=2$ supersymmetry.

\newsec{ ``Large" Flux}

In the study of flux vacua it is important to find out how many distinct fluxes there are
(not related  by duality)  which give rise to allowed vacua. 
In particular one would like to know if this number is finite or infinite. 
A related question is to ask if the flux can be made ``large" subject to the restriction that the 
total D3-brane charge is fixed. In this section we 
 examine this question for the $K3\times T^2/Z_2$ case. We construct a  one parameter 
family of fluxes which are inequivalent, all of which have the same contribution to the 
 D3 brane charge, \defnflux. 
However only one of these sets of fluxes gives rise to a vacuum, for all the other values
of the parameter one can show that there is no susy preserving or susy breaking vacuum.

The idea behind the construction is as follows. 
Consider starting with a case where the  flux vectors
$\alpha_x,\alpha_y,\beta_x,\beta_y,$ yield a consistent susy solution.
Suppose a null vector, $v \in H^{2}(K3,{\bf Z})$, $v \cdot v=0$, exists
which is orthogonal to all four of these flux vectors.
Then one can consider modifying the flux vectors  by adding arbitrary
(even integer) multiples of the null vector $v$. E.g
$\alpha_x \rightarrow \alpha_x + n_x v$ etc.
As a concrete example consider starting with the  flux, \fluxs,  with $dim V_{flux}=2$, discussed
in \S4.1. A null vector $v$ can always be found in this case orthogonal to $V_{flux}$.
We can now modify the flux vectors as follows, $\alpha_y,\beta_x,\beta_y$ are unchanged,
while,
\eqn\modalx{\alpha_x \rightarrow \alpha_x + 2 n v.}
It is quite straightforward to show that the resulting family consists of distinct fluxes
not related by $U$ duality transformations. 
E.g. large coordinate transformations on K3 cannot turn the
starting flux, to \modalx. This is because to begin,   $\alpha_x=\beta_y,$ but only
 $\alpha_x$
 varies as $n$ is changed.  Similarly $S$ duality which exchanges ${\cal H}_3$ and $F_3$
and T-duality on the $T^2$ also do not relate these different choices.
Thus we see that as $n$ is varied we have a one parameter family of different 
fluxes with the same value of 
$N_{flux}$. 

Let us now examine if the modified fluxes  lead to allowed vacua.
For a susy preserving vacuum the flux must be of type $(2,1)$ and primitive. 
Clearly the equations \condc,\condd,\conde, still continue to hold for the same values of
$\phi,\tau$ as before and the inequality \condf\ is still met.
So with $\tau$  and $\phi$ fixed at the same value as before
the $(2,1)$ condition is met for the new fluxes as well.
Note that the new complex structure of  $K3$ will change on modifying the flux vectors.
{}From \cchoice\ we see that $G_{\bar z}$ and therefore $\Omega$ will be different.

Next let us consider the primitivity condition. It is easy to see that this  cannot be met 
for the modified flux vectors. This  follows from the discussion in \S3.3.   
 The inner product matrix in $V_{flux}$, after the modification, will have one null eigenvector.
As a result, one cannot find a spacelike K\"ahler two-form in $K3$ orthogonal to $V_{flux}$.

So we see that  for $n\ne 0$ there is no susy preserving solution. 

Next let us ask about susy breaking vacua. In this case the flux can be a sum   
of type 
$(2,1)$ and primitive,  $(0,3)$ and  $(1,2)$. The $(1,2)$ term must be of the type,
$J\wedge \alpha$, where $J$ is the K\"ahler two form and $\alpha$ is a non-trivial
one form of type $(0,1)$. 
It is easy to see that this implies that, $G_z$, is of type, $(1,1)$,
and satisfies the condition, $G_z \cdot J=0$,  while, $G_{\bar z}$ can be expressed as, 
$G_{\bar z}=c_1 \Omega + c_2 {\bar \Omega}  + c_3 J$. 
Note that,  $\Omega \cdot J={\bar \Omega} \cdot J=0$, so that $\Omega, J $ 
together define a spacelike
three plane.
As a result, the  conditions, \condc, \condd,   must still hold. Together with the condition, 
$J \cdot G_z$, these
 imply that the real and imaginary parts of $G_z$ must be time like vectors.
But this condition cannot be met for a non-zero $G_z$ since  $V_{flux}$ is spanned 
 by two spacelike vectors, \fluxs, and one null vector, $v$.
Finally, it is also easy to show that for the modified flux, \modalx, and nonsingular 
values of $\phi,\tau$, $G_z$ cannot vanish. Thus we conclude that for $n \ne 0$
there are no susy breaking solution. 

In summary we have constructed a one parameter family of fluxes in this section, 
all of which correspond to the same value of $N_{flux}$, \defnflux. 
However, only one of them leads to an allowed vacuum.

\newsec{Duality}

\def\ah{\hat\alpha}
\def\bh{\hat\beta}
\def\gh{\hat\gamma}
\def\detjht{{\rm det}_{xy}{\hat j}}
\def\dtjxy{{\rm det}_{xy}j}

In this section we will study various dual descriptions of the IIB theory on 
$K3 \times T^2/Z_2$ in the presence of flux. 

One T-duality will take us to Type IIA (or Type $I^{'}$) with $O8$ planes. 
Two T-dualities will lead to a Type I description. Finally a further 
S-duality will give rise to a Heterotic  description. 
The dual descriptions are  not  (conformally) Calabi-Yau spaces,
in fact they are  not even K\"ahler manifolds.
They are related to compactifications  of the Heterotic string with Torsion,
\refs{\andytors, \keshav} and the more recent constructions in 
\refs{\janlouis \dlust {--} \kstt}.
A general understanding of such compactifications is still not available in the literature.
Our discussion will parallel that of \kstt, and we will use similar notation below. 

The supergravity backgrounds for the 
three duals mentioned above can be explicitly 
constructed  for all IIB flux compactifications on $K3\times T^2/Z_2$.
The ${\cal H}_3$ flux in the starting  theory must have two legs along the $K3$ and one along the 
$T^2$, see \deffluxh. The isometries along both the directions of the $T^2$ can  be
then made manifest by   choosing a gauge where the two-form NS gauge 
potential, $B$, has no dependence on the two $T^2$ directions. One and two
T-dualities along these directions can then be explicitly carried out
and the supergravity backgrounds can be obtained using \refs{\rBUSH \HassanBV
{--} \BHO}.
Following this it is straightforward to carry out the S-duality as well. 

However, since several moduli are fixed in the starting description, it is not always 
 possible to go to a region of moduli space where the sugra description is valid 
in the dual theory. This problem can be avoided in cases where the moduli are only partially
lifted,  in such situations  the dual sugra description can sometimes be  reliable. 
An example of such a compactification for the $T^6/Z_2$ case was explored in \kstt,   
similar examples for $K3\times T^2$ can also be constructed, but we will not elaborate on 
them here.

We will also construct a superpotential in the dual theories below. It will involve the
appropriate RR and NS fluxes  as well as certain ``twists" in the geometry.

One comment about notation.  
In this section $\mu,\nu,...$ denotes all the compact directions.
The two directions of the $T^2$ are denoted by $x,y$. We will carry out  T-duality along the 
$x$ direction first and then along the $y$  direction.
$\alpha,\beta,...$ will   denote compact directions other
than $x$ and $\ah,\bh,...$  compact directions other than $x$ and $y$.

\subsec{One T-duality}

We  carry out the  T-duality along the  $x$ direction of the $T^2$.
As was mentioned above one can always choose a gauge in which ${\cal B}_{x {\hat \alpha}}$
is independent of the $T^2$ directions. 

We denote  the metric
of IIB theory before duality by  $j_{\mu\nu}$, and the metric in the IIA theory
after T-duality by $g_{\mu\nu}$.
The metric of the resulting 
manifold ${\cal M}'$ is given by
\eqn\iiametric{
ds^2_{{\cal M}'} = {1\over j_{xx}} \eta^x \eta^x 
+ {1\over j_{xx}} \left(\dtjxy\right) dy^2
+ j_{\ah\bh} dx^{\ah}dx^{\bh}
}
where the one form $\eta^x = dx - {\cal B}_{x\ah}dx^{\ah}$, which can 
also be written as $\eta^x = g_{x\mu}dx^{\mu}/g_{xx}$ and 
$\dtjxy~=~j_{xx} j_{yy} - j_{xy}^2~.$ 
Note that ${\cal B}_{x\ah}dx^{\ah}$ varies non-trivially along the $K3$.
As a result the resulting compactification is a sort of ``twisted" analogue
of the  $K3 \times T^2$  space \foot{It can be shown by an explicit calculation
that due to the non-trivial twist, ${\cal M}'$ is not Ricci flat. It follows then
that it cannot be a Calabi Yau manifold.}. It would be quite useful to have a more complete
understanding of such compactifications. In \kstt, it was shown that the dual 
compactifications could be thought of as cosets which are  generalisations of the 
nilmanifold. It would  be interesting to ask if there is a similar description in 
the present case. 

Two more comments are in order at this stage. 
First, besides the metric, the RR forms $F_4,F_2,$ and the NS form $H_3,$ are also excited
in this background. Their values can be determined using the formulae in App.C,  
\HassanBV, \BHO, but we will not do so here. 
Second, one can define a 
 two form 
\eqn\defspind{\omega_{(x)} = - d(g_{x\alpha}dx^{\alpha}/g_{xx})~.}
This is the $x$ component of the antisymmetrised spin connection.
The $x$ direction is an isometry of the IIA metric  and  $\omega_{(x)}$
is the field strength of the Kaluza Klein gauge symmetry  associated with this isometry. 
It will enter our discussion of the superpotential below. 

{\it Superpotential}

In writing down a superpotential in the dual theory which is the analogue of 
\superpot, it is first convenient to define an almost complex structure (ACS) as follows. 

Define the one-form 
\eqn\hmorph{
\eta^z = \eta^x + i \sqrt{det_{xy}j} dy~.
}
The metric \iiametric\ can then  be written as
\eqn\hermmet{
ds^2_{{\cal M}'} = {1\over j_{xx}} \eta^z{\bar\eta}^{\bar z} + ds^2_{K3} ~,}
with $ds^2_{K3}$ denoting the metric over $K3$,
\eqn\metsup{ ds^2_{K3} = j_{\ah\bh} dx^{\ah}dx^{\bh} ~.}

Consider a complex structure on $K3$ compatible with the metric, \metsup.
Let $dz^1,dz^2$ be holomorphic one forms (in the space spanned 
by $dx^{\ah}, \ah, 1, \cdots 4$) with respect to this complex structure.
Then the required  almost complex structure we use below is defined by specifying a basis 
of holomorphic one forms to be $\eta^z, dz^1,dz^2$. 
A holomorphic $(3,0)$ form $\Omega_{IIA}$ can be constructed, it is 
\eqn\choiceacs{
\Omega_{IIA} = \Omega\wedge\eta^z ~,
}
where $\Omega \sim dz^1 \wedge dz^2$ is  the holomorphic $(2,0)$ form on $K3$.
This ACS  is analogous to that used in \kstt.  While we omit the details here,
the spinor conditions take a convenient
form with this choice of  ACS and as a result  a superpotential can also  be easily  
constructed.

The  superpotential is   given by 
\eqn\iiaspot{
W_{IIA} = \int_{\cal M'} G_{IIA} \wedge \Omega_{IIA}
}
where 
\eqn\giia{
G_{IIA} = \left( \tilde{F}_{4(x)}
+ g_{xx} \eta^x\wedge F_2 \right) - i \bigl(\sqrt{g_{xx}}/g_s^{\rm
IIA}\bigr) \left( H_3 - g_{xx} \eta^x \wedge \omega_{(x)} \right)~.}

Here we have used the definitions 
\eqn\foriia{\eqalign{
 {\tilde F}_4 = dC_3 + A_1 \wedge H_3 ~,~ 
\left[{\tilde F}_{4(x)}\right]_{\alpha\beta\gamma}
={\tilde F}_{x\alpha\beta\gamma}~.
}} 
The last term of the superpotential 
contains a component of the  spin connection which was discussed above in \defspind. 
It arises from the term in the starting IIB superpotential \superpot,
proportional to ${\cal H}_3$ with one leg along the $x$ direction. 

Evidence in support for this superpotential includes the  following.
First, susy requires that $G_{IIA}$ is of type $(2,1)$ with respect to the 
ACS defined above. This condition is obtained by varying the superpotential \iiaspot.
Second, the various terms in the superpotential \iiaspot, correctly account for the 
tension of various BPS domain walls in the theory. In particular the last term,
involving the spin connection, gives the tension of a KK monopole related to the 
$x$ isometry direction. 
 
\subsec{Two T-dualities}
We can now further T-dualise along the $y$ direction to obtain 
a type I compactification with $F_3$ flux and twists in the geometry.
We denote the dual manifold as ${\cal M}$. The metric on ${\cal M}$ is 
\eqn\txymetric{
ds^2_{\cal M} = \left({\hat j}_{xx} {\hat\eta}^x{\hat\eta}^x 
     + {\hat j}_{yy} {\hat\eta}^y{\hat\eta}^y 
     + 2 {\hat j}_{xy} {\hat\eta}^x{\hat\eta}^y \right)
     + ds^2_{K3} 
}
with
\eqn\defjxy{
{\hat j}_{xx} = {j_{yy}\over\dtjxy}  ~,~
{\hat j}_{yy} = {j_{xx}\over\dtjxy}  ~,~
{\hat j}_{xy} = - {j_{xy}\over\dtjxy}, 
}
and 
\eqn\etax{\eqalign{
\hat\eta^x = dx + {{\hat j}_{yy}\over \detjht}
        \left( {\hat j}_{xx} {\hat j}_{(x)}
       - {\hat j}_{xy} {\hat j}_{(y)} \right),}}

\eqn\etay{\eqalign{
\hat\eta^y = dy + {{\hat j}_{xx}\over \detjht}
         \left( {\hat j}_{yy} {\hat j}_{(y)}
       - {\hat j}_{xy} {\hat j}_{(x)} \right)}}
Note, we use 
` ${\hat{}}$ ' to  denote quantities in the
type I theory.

A superpotential can be defined in this case as well. 
The metric \txymetric\ can be rewritten as
\eqn\dscalm{
ds^2_{\cal M} = {\hat j}_{xx} {\hat\eta}^z {\hat{\bar\eta}}^{\bar z}
             + ds^2_{K3}}
An almost complex structure can be now be  specified by defining one  
holomorphic one-form to be 
$${\hat\eta}^z = {\hat\eta}^x + {\hat\tau}{\hat\eta}^y ~,$$
and  two additional holomorphic one-forms to be compatible with the complex structure 
of $K3$. 

A $(3,0)$ form ${\hat \Omega}$ is then defined by 
\eqn\twothol{
{\hat\Omega} = \Omega\wedge{\hat\eta}^z ~.
}
 where $\Omega$ is the holomorphic two-form of $K3$.

The resulting superpotential is  
\eqn\typIsup{
{\hat W} = \int_{\cal M} {\hat G} \wedge {\hat\Omega}
}
where 
\eqn\ghat{\eqalign{
{\hat G} = & \left( {\hat j}_{xx} \hat\eta^x
          + {\hat j}_{xy} \hat\eta^y\right) \wedge {\hat F}_{3(y)}
          - \left( {\hat j}_{xy} \hat\eta^x + {\hat j}_{yy} \hat\eta^y\right) 
          \wedge {\hat F}_{3(x)} \cr
& -  \left(i \over g_s^I\right) \sqrt{\detjht} \left( 
{\hat j}_{xx} \hat\eta^x \wedge d{\hat j}_{(x)}
+ {\hat j}_{yy} \hat\eta^y \wedge d{\hat j}_{(y)} \right)
}}
Some additional   notation used in the above formula is as follows.
The one and two forms 
${\hat j}_{(x)},~ {\hat{ H}}_{3(x)},~{\hat F}_{3(x)}$ 
are given as
\eqn\exps{
{\hat j}_{(x)} = {1\over {\hat j}_{xx}} {\hat j}_{x\ah} dx^{\ah} ~,~
 \left[{\hat F}_{3(x)}\right]_{\ah\bh}
= \left[{\hat F}_3\right]_{x\ah\bh} ~,~
}
and similar expressions for the quantities carrying the index `$y$'.
Also we have used the definition
\eqn\fordtj{ \detjht = \left( {\hat j}_{xx} {\hat j}_{yy} 
           - {\hat j}_{xy}^2 \right), }
and the type I string coupling $g_s^I = e^{\phi^I}$ \foot{ 
Note that 
The general expression for the superpotential in \kstt\
after T-dualizing along both $x$ and $y$ appears with terms
containing ${\hat{\tilde F}}_{5(xy)}~,~{\hat F}_1$ and ${\hat H}_3$
but are absent here in the superpotential \typIsup\ . This is 
due to the fact that terms containing the above quantities arise
from $\left[F_3\right]_{\ah\bh\gh},\left[F_3\right]_{xy\ah},
\left[H_3\right]_{\ah\bh\gh}$ of the original type IIB theory. 
However they are projected out in $K3\times T^2/Z_2$
compactification of IIB.}.

\subsec{Heterotic Dual}
Making a further S-duality we obtain heterotic theory on a manifold 
${\cal M}_{\rm het}$ whose metric  (in string frame) is given by
\eqn\hetdssqr{
ds^2_{\rm het} = j^h_{xx} \eta_h^z{\bar\eta}_h^{\bar z} 
+  g_s^{\rm het} ds^2_{K3} ~.
}
We denote the metric components in the heterotic theory with a superscript
` $h$ ' , which are related to the type~I metric by a factor of the string
coupling $g_s^{\rm het}$ :
$${\hat j}_{\mu\nu} = {1\over g_s^{\rm het}} j^h_{\mu\nu} ~.$$

We choose an almost complex structure in the heterotic theory,
which agrees with the one described above for the Type I case. 
The corresponding $(3,0)$ form is then given by ${\hat \Omega}$, \twothol,
and the superpotential is the same as \typIsup. 
Expressed in heterotic language this takes the form,   
\eqn\hetsupt{
W_{\rm het} = \int_{{\cal M}_{\rm het}} G_{\rm het} \wedge {\hat \Omega} 
}
with 
\eqn\ghetexp{\eqalign{
G_{\rm het} = & \left(j^h_{xx} {\hat \eta}^x 
             + j^h_{xy} {\hat \eta}^y\right)\wedge H_{3(y)}
   - \left(j^h_{xy} {\hat \eta}^x + j^h_{yy} {\hat \eta}^y\right)\wedge  H_{3(x)} \cr
& -  i \sqrt{{\rm det}_{xy}j^h} \left(
j^h_{xx} {\hat \eta}^x \wedge dj^h_{(x)}
+ j^h_{yy} {\hat \eta}^y \wedge dj^h_{(y)} \right).
}}

The two forms $H_{3(x)},H_{3(y)}$, above are 
given by
\eqn\hthrxy{
\left[H_{3(x)}\right]_{\ah\bh} = \left[H_3\right]_{x\ah\bh} ~,~
\left[H_{3(y)}\right]_{\ah\bh} = \left[H_3\right]_{y\ah\bh} ~.
}

Before closing let us note that heterotic compactifications with $H_3$ flux
have been considered in \refs{\haridass,\andytors,\keshav,\bbdg,\dlust}.
The ACS we have defined above is not integrable in general. However,
the heterotic compactifications, and also the Type I models of the preceding section,
are in fact complex manifolds, as can be shown from the analysis in \andytors,
and admit an integrable ACS. It would be interesting to ask what form the superpotential
takes in terms of this complex structure.

\newsec{The $T^6/Z_2$ orientifold}

In this section we will discuss two aspects of  flux compactifications on
$T^6/Z_2$. \S8.1\ presents   general susy preserving  solutions for the $T^6/Z_2$
compactification, in analogy with the discussion in \S3\ for $K3 \times T^2/Z_2$. 
\S8.2\ deals with a family of susy breaking solutions, similar to the one
in \S6\ with complex structure stabilised at extreme values for large flux. 

\subsec{General Susy Solutions}
Here we discuss general susy preserving solutions in the $T^6/Z_2$ model.
We will build on the discussion in \kst. 
The essential idea is similar to \S3\ above.
The requirements on the flux for the existence  of susy solutions can be stated
in terms of simultaneous solutions to two polynomial equations. 
Once these requirements are met   the complex structure moduli can be determined
in terms of the flux.

We start with the  superpotential,
\eqn\kstsup{
W = \int G_{3} \wedge \Omega_{3}.
}
We will use the notation of \kst, below. In particular,
\eqn\fthrhthr{\eqalign{
& {1\over (2\pi)^2\alpha'} F_3 
= a^0 \alpha_0 + a^{ij} \alpha_{ij} + b_{ij} \beta^{ij} + b_0 \beta^0 \cr
& {1\over (2\pi)^2\alpha'} {\cal H}_3 
= c^0 \alpha_0 + c^{ij} \alpha_{ij} + d_{ij} \beta^{ij} + d_0 \beta^0, 
}}
where the three forms $\alpha_{ij}$ and $\beta^{ij}$ are defined in \kst.
Susy solutions satisfy the conditions $W=\partial_iW=0$, where $i$ denotes all 
complex structure moduli. 
It was shown in Sec. \S 4.4 of \kst,  that
the equation $\partial_{\tau^{ij}}W = 0$ can be used to solve for $\tau^{ij}$
in terms of $\phi$ as
\eqn\slvfrtij{
\left(\cof(\tau - \tilde{A})\right)_{ij} = 
  \left(\cof{\tilde{A}}\right)_{ij} + {\tilde B}_{ij}
}
where we define $A^0 = a^0 - \phi c^0 ~,~ A^{ij} = a^{ij} - \phi c^{ij}$
and ${\tilde A}^{ij} = A^{ij}/A^0$ and similar definitions for $B_0,B_{ij}$
and ${\tilde B}_{ij}$ . Solving \slvfrtij\ for $\tau^{ij}$ we get
\eqn\slvdtij{
\tau^{ij} = {\tilde A}^{ij} 
           + { \left( \cof\mu\right)^{ij} \over \sqrt{\det(\mu)}}
}
where we define $\mu_{ij}$ as 
\eqn\dfnmij{
\mu_{ij} = \left(\cof{\tilde{A}}\right)_{ij} + {\tilde B}_{ij}
}
Thus once we know $\phi$ for a given flux we can determine $\tau^{ij}$ .
The conditions for $\phi$ are obtained from the remaining two 
equations $W = 0$ and $\partial_{\phi}W =0$. Combining $W=0$ and 
$\partial_{\tau^{ij}}W = 0$ gives a quartic for $\phi$
\eqn\fsteqn{
B_0 {\rm det}A - A^0 {\rm det}B + (\cof A)_{ij} (\cof B)^{ij} 
+ {1\over 4} (A^0 B_0 + A^{ij} B_{ij})^2 = 0
}
The derivation of this equation is given in App. B of \kst. In addition we need
$\partial_{\phi}W$ to be zero, which combined with $W=0$ gives
$$ a^0 \det({\tau}) - a^{ij} (\cof \tau)_{ij} - b_{ij} \tau^{ij} - b_0 = 0$$
Eliminating $\tau^{ij}$ from above we get
\eqn\elmntau{ \Delta + {1\over \sqrt{\det\mu}} (\cof\mu)^{ij}\Sigma_{ij} = 0
}
where $\Sigma_{ij}$ and $\Delta$ are defined as 
\eqn\sgmij{
\Sigma_{ij} \equiv {1\over 3} a^0 \left(\cof{\tilde A}\right)_{ij}
                   + {1\over 3} a^0 {\tilde B}_{ij}
                   - \epsilon_{ikm} \epsilon_{jln} a^{kl}{\tilde A}^{mn}
                   - b_{ij}
}
and
\eqn\anddlta{
\Delta \equiv {\tilde A}^{ij} \Sigma_{ij} 
              + {2 \over 3} {\tilde A}^{ij} {\tilde B}_{ij} 
              - a^{ij} {\tilde B}_{ij} - b_0
}
We take the square of \elmntau\ to obtain a polynomial in $\phi$ : 
\eqn\sndeqn{
\left(\Sigma_{ij}(\cof\mu)^{ij}\right)^2 - \det(\mu) \Delta^2 = 0
}
The equations \fsteqn\ and \sndeqn\ can be rewritten as 
\eqn\tsixfst{
{\cal F}_4(\phi) \equiv {\hat f}_0 + {\hat f}_1 \phi + {\hat f}_2 \phi^2 
         + {\hat f}_3 \phi^3 + {\hat f}_4 \phi^4 = 0
}
and
\eqn\tsixsnd{
{\cal G}_{12}(\phi) \equiv {\hat g}_0 + {\hat g}_1 \phi 
+ {\hat g}_2 \phi^2 + ... + {\hat g}_{12} \phi^{12} = 0 ~.
}
It is straightforward to obtain the coefficients ${\hat f}_i$ and ${\hat g}_i$. 
They are determined in terms of the integers $a^0, a^{ij}, b_0, b_{ij} \cdots $
appearing in the expressions of $F_3$ and ${\cal H}_3$ as given in \fthrhthr.
However, the expressions are quite lengthy and hence and we will not write 
the precise formulae for them here.  

As before, in order to have a nonsingular solution, the polynomials 
${\cal F}_4(\phi)$ and ${\cal G}_{12}(\phi)$ must admit a common 
quadratic (say ${\cal W}_2(\phi)$) . Thus,
\eqn\tsxfctr{
{\cal F}_4(\phi) = {\cal F}_2(\phi) {\cal W}_2(\phi)~,~
{\cal G}_{12}(\phi) = {\cal G}_{10}(\phi) {\cal W}_2(\phi)
}
where ${\cal F}_2(\phi)$ and ${\cal G}_{10}(\phi)$ are polynomials 
in $\phi$ with coefficients $u_i$ and $v_i$ respectively. Note that 
here the subscripts in ${\cal F} , {\cal G}$ and ${\cal W}$ denotes 
the degree of the polynomial.

The general solution can be stated in terms of a $15\times 14$
matrix ${\hat{\cal M}}$ defined interms of the quantities ${\hat f}_i$ and 
${\hat g}_i$ as
\eqn\defnmt{
{\hat{\cal M}} = \left(
\matrix{ &{\hat f}_0 & 0   & 0   & . & . & . & 0 & 0 
                                   & - {\hat g}_0 &   0   & 0     \cr
         &{\hat f}_1 & {\hat f}_0 & 0   & . & . & . & 0 & 0 
                            & - {\hat g}_1 & - {\hat g}_0 & 0     \cr
         &{\hat f}_2 & {\hat f}_1 & {\hat f}_0 & . & . & . & 0 & 0 
                     & - {\hat g}_2 & - {\hat g}_1 & - {\hat g}_0 \cr
         & .  & .   & .   &  & &    & .  & . &   .   &   .   &   .   \cr
         & .  & .   & .   &  & &   & .  & . &   .   &   .   &   .   \cr
         & .  & .   & .   &  & &   & .  & . &   .   &   .   &   .   \cr
        & 0  & 0   & 0   & . & . & .  & {\hat f}_3 & {\hat f}_2 
               & - {\hat g}_{12} & - {\hat g}_{11} & - {\hat g}_{10} \cr 
         & 0  & 0   & 0   & . & . & .  & {\hat f}_4 & {\hat f}_3 
                         &  0    & - {\hat g}_{12} & - {\hat g}_{11} \cr
         & 0  & 0   & 0   & . & . & . & 0 & {\hat f}_4 
                         &  0    & 0       & - {\hat g}_{12}
}
\right)
}
and a column vector ${\cal X}$ defined as
\eqn\clmnt{
{\cal X} = \left( \matrix{ & u_0 \cr & u_1 \cr &  u_2 \cr & v_0 \cr
& v_1 \cr & . \cr & . \cr & .  \cr & v_{12}} \right)
}
The column vector must satisfy
\eqn\tsixvec{
{\hat{\cal M}}{\cal X} = 0
}
In addition the solution obtained for $\phi$ must be complex. The coefficients 
${\hat w}_i$ in the polynomial ${\cal W}_2(\phi)$ can be obtained from 
\tsxfctr\ in terms of $u_i$ and $v_i$, which themselves are solved in terms of 
${\hat f}_i ~,~ {\hat g}_i$ from \clmnt\ . They must obey
$$ {\hat w}_1^2 < 4 {\hat w}_0 {\hat w}_2 $$
In addition the imaginary part of $\tau_{ij}$ as given in \slvdtij\ also
must be non zero.

\subsec{Large Flux on Tori} 
Here we construct a one parameter family of flux for the $T^6/Z_2$ case 
 analogous to the one discussed in the \S6 for $K3 \times T^2/Z_2$. 
The family consists of  fluxes unrelated by duality, but with a fixed,  $N_{flux}$, \defnflux.
As in \S6, we find there is an allowed vacuum for only one value of the parameter.

 We will consider tori of the form 
$T^4 \times T^2$ and turn on three-form flux with two legs along the $T^4$ and one 
leg along the $T^2$ (this is consistent with the $Z_2$ orientifolding). 
The discussion of \S6\ can now be largely carried over to this case with
the  $T^4$ replacing 
$K3$.

As a concrete example,  we consider
the case where the three-flux takes the form
\eqn\sbta{\eqalign{{1\over (2\pi)^2\alpha'}
F_3 &= 2 e_1 \wedge dx + 2 e_2 \wedge dy \cr
{1\over (2\pi)^2\alpha'} {\cal H}_3 & = -2 e_2 \wedge dx + 2 e_1 \wedge dy, }}
where in our notation $0\le x,y\le 1$ are coordinate on the $T^2$ and 
$e_1,e_2, \cdots$ are two -forms on  $T^4$ as discussed in App.A. 
In the notation of \S3, \S6, this corresponds to taking,
$\alpha_x=\beta_y = 2 e_1$, and, $\alpha_y=-\beta_x= 2 e_2$. 

It is straightforward to show that in this case a susy preserving solution exists 
where $\phi=\tau=i$, with $\tau$ being the complex structure of the $T^2$,
and, where the $T^4=T^2 \times T^2$ with the complex structure of both  $T^2$'s
being stabilised at the same point in moduli space, 
$\tau^1=\tau^2=i$. The Primitivity condition can also then be easily met by a
 K\"ahler two-form,
\eqn\sbtkah{J=\sum_ig_{i{\bar i}}dz^idz^{\bar i},} 
where $i=1, \cdots 3,$ refers  to the three two-tori respectively .

Now consider the vector 
\eqn\sbtn{v=e_3 + e_4,}
valued in $H^2(T^4,{\bf Z})$ (again we refer the Reader to App.A for definitions). 
It is null, i.e., $v \cdot v=0$, and orthogonal to $e_1,e_2$. 

We can now modify the flux vectors as follows. Keep, $\alpha_y,\beta_x,\beta_y,$ the same
and take 
\eqn\sbtmodv{\alpha_x \rightarrow \alpha_x + 2 n v.}
The $(2,1)$ condition can then be met if $\phi=\tau=i$ and 
$\Omega$ (the holomorphic two-form of $T^4$ ) meets the condition 
\eqn\sbtomega{\Omega=(\alpha_x -\phi \beta_x) \tau - (\alpha_y -\phi \beta_y).}
The primitivity condition however cannot be satisfied. 
As a result no susy preserving solution exists for the modified flux, \sbtmodv.
An argument quite similar to the one in \S6 also  shows that no supersymmetry breaking 
solution exists. Hence we conclude that for the modified flux, \sbtmodv, there are are  
no allowed vacua.

\newsec{Discussion}

In this paper we have discussed flux  compactifications of IIB string theory.
Our emphasis was on the $K3 \times T^2/Z_2$ compactification, we also discussed some aspects of
the $T^6/Z_2$ case. 

There are several open questions which remain.

The $K3 \times T^2/Z_2$ compactification has  $D7$-branes present in it.
Our analysis   did not consider the 
effects of exciting the gauge fields on these seven branes.
It would be interesting  to do so, both in  the supersymmetry conditions and the resulting 
superpotential. We mentioned in \S2.4, that generically one expects the moduli associated
with the locations of the D7-branes on the $T^2$ to be lifted, once flux is turned on.   
With gauge fields excited one expects this to continue to be true. 
 For example, the D7-branes can acquire D5-brane charge, if the gauge field which is 
excited  has non-trivial first Chern class. In the presence of  a magnetic $F_3$ flux  
the seven branes will then experience a force along the $T^2$ directions.

We formulated in our discussion above the conditions which must be satisfied by the flux  for 
susy preserving solutions to exist. It would be interesting to determine how many distinct 
fluxes (unrelated by duality) there are which meet these conditions. In particular,
one would like
to know if this number is finite or infinite and if the  fluxes can be made large subject to 
the restriction of total D3-brane charge being held fixed.
We saw above by constructing explicit examples
 that this restriction does allow for an  
 infinite number of distinct fluxes, and therefore to large  flux.
However, even allowing for susy breaking we found that only one set of fluxes, in the one 
parameter family we constructed, gave rise to a 
stable vacuum. In fact, so far,  by varying the flux, we have been unable to construct am 
infinite family 
of vacua with broken or unbroken susy.  It will be useful to settle this issue conclusively in the
future. This will be a useful step in addressing the question of how may ${\cal N}=1$ vacua there are in string theory.

The next logical step in the study of flux vacua, in continuation of  \kst\ and this paper,
 would be to consider (orientifolds of)
Calabi-Yau threefolds with flux. At the moment, we do not see how to directly   generalise
 the  techniques devised  
for $K3$ to Calabi Yau threefolds.  Perhaps, the best approach might be to 
consider a simple case with few complex structure moduli and explicitly evaluate the 
superpotential \foot{We thank S. Kachru for preliminary discussions on this.}.

We examined some dual theories related to the flux compactifications of 
IIB string theory above, and saw that they are not Calabi-Yau spaces. We also obtained a 
superpotential in these dual descriptions.  Much more can be done   
along this direction. For example, one would like to construct examples which cannot be related to 
Calabi-Yau compactifications via duality. 

Finally, from the viewpoint of moduli stabilisation, the most serious  limitation of these models 
is that the volume modulus (in the IIB description) is not stabilised. 
It would be illuminating to consider various additional effects which could lift this
direction. Non-perturbative gauge dynamics on the world volume of D7-branes, present in
the $K3 \times T^2/Z_2$ example   considered above, might provide a tractable example. 

\medskip
\centerline{\bf{Acknowledgements}}
\medskip
It is a great pleasure to thank S. Kachru for extensive discussions. 
We are also grateful to K. Dasgupta for discussion  and for sharing results 
prior to publication.
We also thank A. Dabholkar, M. Douglas, and N. Fakhruddin, J. Maldacena,  S. Mukhi, 
G. V. Ravindra, M. Schulz and A. Sen for their comments.  
S.P.T. acknowledges support from the Swarnajayanti Fellowship,
Department of Science and Technology, Government of India. The work of P.K.T.
and S.P.T. was supported by the DAE, Government of India. Most of all,
we  thank the people of India for enthusiastically  supporting research 
in  string theory.

\appendix{A}{$\Gamma^{3,19}$ Lattice}

In this appendix we give some details about the integral cohomology of $K3$. Our conventions for orientation etc on $K3 \times T^2/Z_2$ are also explained. Towards the end 
we discuss the four-torus, $T^4$, this is relevant to the discussion in 
\S8.2.

The integral cohomology $H^2(K3,{\bf Z})$ has the  structure of an even self-dual
lattice of $\Gamma^{3,19}$ of signature $(3,19)$. 
We can choose a basis $\{e_i\}$ for $\Gamma^{3,19}$ such that the inner
products of the basis vectors {\foot{We denote the inner product of lattice
vectors by a {\it dot} {i.e.} $(\alpha,\beta) = \alpha\cdot\beta$.}}
\eqn\ktmetric{
g_{ij} = (e_i,e_j) = \int_{K3} e_i \wedge e_j
}
is given by the following matrix
\eqn\gijexp{ g_{ij} = \left(\matrix{ & H_{3,3} & 0 & 0 \cr
                                     & 0 & - {\cal E}_8 & 0 \cr
                                     & 0 & 0 & - {\cal E}_8
}
\right)}
where the matrix $H_{3,3}$ is defined as
\eqn\hthrethre{
H_{3,3} = \left(\matrix{
&0 &1 &0 &0 &0 &0 \cr
&1 &0 &0 &0 &0 &0 \cr
&0 &0 &0 &1 &0 &0 \cr
&0 &0 &1 &0 &0 &0 \cr
&0 &0 &0 &0 &0 &1 \cr
&0 &0 &0 &0 &1 &0 }
\right)
}
 and ${\cal E}_8$ is the
Catran matrix of $E_8$ algebra :
\eqn\cartan{
{\cal E}_8 = \left(\matrix{
&2&\!\!\!\!-1& 0& 0& 0& 0& 0& 0 \cr
&\!\!\!\!-1& 2 &\!\!\!\!-1& 0& 0& 0& 0& 0 \cr
&0&\!\!\!\!-1& 2 &\!\!\!\!-1& 0& 0& 0&\!\!\!\!-1 \cr
&0& 0&\!\!\!\!-1& 2 &\!\!\!\!-1& 0& 0& 0 \cr
&0& 0& 0&\!\!\!\!-1& 2 &\!\!\!\!-1& 0& 0 \cr
& 0& 0& 0& 0&\!\!\!\!-1& 2 &\!\!\!\!-1& 0 \cr
&0& 0& 0& 0& 0&\!\!\!\!-1& 2& 0\cr
& 0& 0&\!\!\!\!-1& 0& 0& 0& 0& 2
}\right) \ .
}
The basis vectors $(e_1 \ , \ e_2 \ , \ \cdots \ e_6)$ span a
subspace of $H^2(K3,{\bf Z})$ (which we denote as $\Gamma^{3,3}$)
with the metric given by $H_{3,3}$. 
Note that we can make a change of basis such that $H_{3,3} = 2 \eta_{3,3}$
with $\eta_{3,3} = {\rm diag}(1,1,1,-1,-1,-1)$ where as ${\cal E}_8$ remains
the same.
 
Let $\gamma_2$ be an element of the integral homology $H_2(K3,{\bf Z})$.
Integrating an arbitrary two form $\alpha_2 \in H^2(K3,{\bf Z})$ 
over $\gamma_2$ results in an integer. In particular we have
\eqn\diracq{
\int_{\gamma_2} e_i \in {\bf Z} ~.
} 
We turn on $H_3$ and $F_3$ fluxes over three cycles which are of the
type $\gamma_2 \times \gamma_1$ where $\gamma_2$ is defined earlier
and $\gamma_1 \in H_1(T^2/Z_2,{\bf Z})$. Integrating $e_i \wedge dx$
and $e_i \wedge dy$ over $\gamma_2 \times \gamma_1$ results in integers
if $\gamma_1$ is a `full cycle' of $T^2$. However if $\gamma_1$ is
a `half cycle' of $T^2$ (a cycle which is closed in $T^2/Z_2$ but 
not in $T^2$) then the result is a  half integers. 
It was pointed out by Frey and Polchinski \joefrey\ that in order to satisfy 
the Dirac quantization conditions in these cases one needs to turn 
on fluxes due to exotic orientifold planes. As in \kst, here we avoid 
these complications  by choosing the fluxes  
corresponding to the lattice vectors with even coefficients in 
$\Gamma^{3,19}$.

It is helpful to describe the   cohomology basis above in detail in  the $T^4/Z_2$
limit of $K3$. The $E_8 \times E_8$ lattice vectors correspond to the 
$16$ blow up modes of the orbifold. 
Choosing coordinates $x^i,y^i, 0\le x^i,y^i \le 1, i=1,2,$  for the $T^4$,
a basis of  $Z_2$ invariant two-forms is given by 
\eqn\idfn{\eqalign{
& e_1 = \sqrt{2}\left( dx^1\wedge dx^2 - dy^1\wedge dy^2 \right) ~~
~~ e_2 = \sqrt{2} \left( dx^1\wedge dy^2 - dx^2\wedge dy^1 \right) \cr
& e_3 = \sqrt{2} \left( dx^1\wedge dy^1 + dx^2\wedge dy^2 \right) ~~
~~ e_4 = \sqrt{2} \left( dx^1\wedge dx^2 + dy^1\wedge dy^2 \right) \cr
& e_5 = \sqrt{2} \left( dx^1\wedge dy^2 + dx^2\wedge dy^1 \right) ~~
~~ e_6 = \sqrt{2} \left( dx^1\wedge dy^1 - dx^2\wedge dy^2 \right) ~. }}
With a  choice of normalization,
\eqn\tfourlmt{
\int_{T^4/Z_2} dx^1\wedge dx^2\wedge dy^1\wedge dy^2  = -{1\over 2} ~,}
these obey the conditions,
$$ (e_1,e_1) = (e_2,e_2) = (e_3,e_3) = 2 $$
$$ (e_4,e_4) = (e_5,e_5) = (e_6,e_6) = -2, $$
with all other inner products being zero. 
We see then that $e_1, ...,e_6$ form a basis for $\Gamma^{3,3}$.

For completeness let us also note that in the notation of this paper, 
our choice of orientation on the $T^2$ is given by 
\eqn\orien{
\int_{T^2} dx \wedge dy = -1 ~.
}

Finally we discuss the $T^4$ case. The six one forms $e_1, \cdots e_6$ defined above
(without the $\sqrt{2}$ prefactor in normalisation),
\idfn,  form a basis of 
$H^2(T^4,{\bf Z})$. We can define an inner product in this vector space
analogous to \ktmetric.
Holomorphic coordinates on $T^4$ can be defined  by
\eqn\appacs{z^i = x^i + \tau^i_j y^j, i= 1, 2.}
The complex structure is completely specified by the period matrix $\tau^i_j$. 
The holomorphic two-form $\Omega = \lambda (dz^1 \wedge dz^2)$ ($\lambda$ is 
a constant) can be expressed in terms of the basis $e_1, \cdots e_6$ as follows:
\eqn\appaoex{\eqalign{\Omega=& {1 \over 2} (1- det \tau) e_1 + 
{1 \over 2} (\tau^2_2 +\tau^1_1)e_2 + {1 \over 2} (\tau^2_1 - \tau^1_2) e_3 \cr
&{1\over 2} (1 + det \tau) e_4 + {1 \over 2} (\tau^2_2 - \tau^1_1) e_5 
+ { 1\over 2}  (\tau^2_1 + \tau^1_2) e_6.}}

\appendix{B}{Solving the Quartic and Quintic Polynomials}
In this appendix we discuss in more detail the conditions leading to the 
two polynomials, \qurqui a\ and \qurqui b, having a common quadratic factor. 

The polynomials are given by
\eqna\polys
$$ P(\phi) \equiv p_1 \phi^5 + p_2 \phi^4 + p_3 \phi^3
                             + p_4 \phi^2 + p_5 \phi + p_6 = 0 \eqno\polys a $$
$$ Q(\phi) \equiv q_1 \phi^4 + q_2 \phi^3 + q_3 \phi^2
                                 + q_4 \phi + q_5 = 0~, \eqno\polys b ~. $$
where the coefficients $p_i$ and $q_i$ are 
\eqn\pexpr{\eqalign{
p_1 = & - \gxxn \bxyn^2 + \bxxn \bxyn \gxyn - \gyyn \bxxn^2 ~, \cr
p_2 = & \axxn \bxyn^2 + \bxxn^2 \ayyn
        - 2 \axyn \bxyn \bxxn + 4 \gxxn \gyyn \bxxn
        - \bxxn \gxyn^2 ~, \cr
p_3 = & - 4 \gxxn^2 \gyyn - 2 \axxn \bxxn \gyyn - 4 \gxxn \bxxn \ayyn
        + 2 \gxxn \axyn \bxyn \cr
      & + \gxxn \gxyn^2
        - \axxn \bxyn \gxyn + 3 \axyn \bxxn \gxyn ~, \cr
p_4 = & -2 \axyn^2 \bxxn + 4 \gxxn^2 \ayyn + 2 \axxn \bxxn \ayyn
        + 4 \axxn \gxxn \gyyn - 4 \gxxn \axyn \gxyn ~, \cr
p_5 = & - \axxn^2 \gyyn + 3 \gxxn \axyn^2 - 4 \axxn \gxxn \ayyn
        + \axxn \axyn \gxyn ~, \cr
p_6 = & \axxn^2 \ayyn - \axxn \axyn^2 ~.
}}
and 
\eqn\qexpr{\eqalign{
&q_1 = \bxyn^2 - \bxxn \byyn ~, \cr
&q_2 = 2 \gxxn \byyn - 2 \bxyn \gxyn + 2 \bxxn \gyyn ~, \cr
&q_3 = - \ayyn \bxxn + 2 \axyn \bxyn - \axxn \byyn
       - 4 \gxxn \gyyn + \gxyn^2 ~, \cr
&q_4 = 2 \ayyn \gxxn + 2 \axxn \gyyn - 2 \axyn \gxyn ~, \cr
&q_5 = \axyn^2 - \axxn \ayyn
}}
Here we have used the notation
\eqn\notn{\eqalign{
 \alpha_{ij} \equiv \alpha_i . \alpha_j ~,
  \beta_{ij} \equiv \beta_i . \beta_j ~,}}
with $i,j = \{x,y\}$, and 
\eqn\notngma{\eqalign{
 \gamma_{xx} \equiv \alpha_x . \beta_x ~,
 \gamma_{yy} \equiv \alpha_y . \beta_y ~, \  {\rm and} \ \
  \gxyn \equiv \ax . \by + \ay . \bx ~.
}}

Assume  the quadratic factor of the form
$$ W(\phi) \equiv w_1 \phi^2 + w_2 \phi + w_3 $$
is the common factor of $ P(\phi)$ and $ Q(\phi)$ .
Then
\eqna\equt
$$ P(\phi) = ( r_1 \phi^3 + r_2 \phi^2 + r_3 \phi + r_4) W(\phi) \eqno\equt a$$
$$ Q(\phi) = (s_1 \phi^2 + s_2 \phi + s_3) W(\phi)  \eqno\equt b $$
for some $r_i$ and $s_i$ .
This gives in particular
\eqn\inpart{
(s_1 \phi^2 + s_2 \phi + s_3) P(\phi) =
( r_1 \phi^3 + r_2 \phi^2 + r_3 \phi + r_4) Q(\phi)
}
Equating the coefficients of $\phi^n$ from both sides we get
\eqn\ary{\eqalign{
& p_1 s_1 - q_1 r_1 = 0 \cr
& ( p_1 s_2 + p_2 s_1) - ( q_1 r_2 + q_2 r_1) = 0 \cr
& ( p_1 s_3 + p_2 s_2 + p_3 s_1) - ( q_1 r_3 + q_2 r_2 + q_3 r_1) = 0 \cr
& ( p_2 s_3 + p_3 s_2 + p_4 s_1)
                         - ( q_1 r_4 + q_2 r_3 + q_3 r_2 + q_4 r_1) = 0 \cr
& ( p_3 s_3 + p_4 s_2 + p_5 s_1)
                         - ( q_2 r_4 + q_3 r_3 + q_4 r_2 + q_5 r_1) = 0 \cr
& ( p_4 s_3 + p_5 s_2 + p_6 s_1) - ( q_3 r_4 + q_4 r_3 + q_5 r_2) = 0 \cr
& ( p_5 s_3 + p_6 s_2) - (q_4 r_4 + q_5 r_3) = 0 \cr
& p_6 s_3 - q_5 r_4 = 0
}}
Define the matrix
\eqn\aryappendixb{
M \equiv  \left( \matrix{& p_1 & 0 & 0 & - q_1 & 0 & 0 & 0 \cr
& p_2 & p_1 & 0 & - q_2 & - q_1 & 0 & 0 \cr
& p_3 & p_2 & p_1 & - q_3 & - q_2 & - q_1 & 0 \cr
& p_4 & p_3 & p_2 & - q_4 & - q_3 & - q_2 & - q_1 \cr
& p_5 & p_4 & p_3 & - q_5 & - q_4 & - q_3 & - q_2 \cr
& p_6 & p_5 & p_4 & 0 & - q_5 & - q_4 & - q_3 \cr
& 0 & p_6 & p_5 & 0 & 0 & - q_5 & - q_4 \cr
& 0 & 0 & p_6 & 0 & 0 & 0 & - q_5
}\right)}
And the column vector
\eqn\defcolumn{X=\left(\matrix{& s_1 & \cr
                               & s_2 & \cr
                               & s_3 & \cr
                               & r_1 & \cr
                               & r_2 & \cr
                               & r_3 & \cr
                               & r_4 & } \right)} 
Then \ary, can be restated as 
\eqn\deteq{ M \cdot X=0}
which is eq. \condcommon. 
Thus the condition for the two polynomials sharing a common quadratic factor is that 
a non-zero vector $X$ exists satisfying \defcolumn. 

 In terms of the components of $X$,  we find, from \equt b, by comparing powers of $\phi$ 
 that $W(\phi)$ is given by \commonquad,\valw. 

\appendix{C}{Duality Maps}

Here we use the notation of \kstt\ . In particular, we define the four form
field strength ${\tilde F}_4 = dC_3 + A_1\wedge F_3$ in IIA theory and the 
five form field strength 
$${\tilde F}_5 = dC_4 
- {1\over 2} C_2 \wedge {\cal H}_3 + {1\over 2} {\cal B}_2 \wedge F_3~.$$
We denote the T dual direction as $x$ and also we define the one forms
\eqn\defjxgx{j_{(x)} = {1\over j_{xx}} j_{x\alpha} dx^{\alpha}~,~
 g_{(x)} = {1\over g_{xx}} g_{x\alpha} dx^{\alpha} } 
and the exterior derivative 
\eqn\defspin{\omega_{(x)} = - dg_{(x)}~. } 
In addition $F_{n(x)}$ denotes
an $n-1$ form whose components are given by:
\eqn\defnx{[F_{n(x)}]_{i_1, \cdots i_{n-1}}=[F_n]_{xi_1 \cdots i_{n-1}}.}

The Neveu-Schwarz fields transform as 
\eqn\twoab{\eqalign{
& g_{xx} = {1\over j_{xx}} \cr
& g_{x\alpha} = - {{\cal B}_{x\alpha}\over j_{xx}} \cr   
& g_{\alpha\beta} = j_{\alpha\beta} 
                 - {1\over j_{xx}} (j_{x\alpha}j_{x\beta}
- {\cal B}_{x\alpha}{\cal B}_{x\beta})\cr
& B_{x\alpha} = - {j_{x\alpha}\over j_{xx}} \cr
& B_{\alpha\beta} = {\cal B}_{\alpha\beta} 
                - {1\over j_{xx}} (j_{x\alpha} {\cal B}_{x\beta}
                - {\cal B}_{x\alpha} j_{x\beta}) \cr
& g_s^{IIA} = {g_s^{IIB}\over{\sqrt{j_{xx}}}}
}}
Here the left hand refers to fields in the IIA theory and the right hand side 
to fields in the IIB theory. 
For the three form field strength, $H_3$ this takes the form,
\eqn\fstrength{\eqalign{
& H_{3(x)} =  dj_{(x)} \cr
& H_3 = {\cal H}_3 - {\cal H}_{3(x)} \wedge j_{(x)} 
                     - {\cal B}_{(x)} \wedge dj_{(x)} 
}}
The Ramond fields transform as
\eqn\rfields{\eqalign{
& F_{2(x)} = F_1  \cr
& F_2 =  {\tilde F}_{3(x)}  - {\cal B}_{(x)} \wedge F_1 \cr
&{\tilde F}_{4(x)} = {\tilde F}_3  - j_{(x)} \wedge {\tilde F}_{3(x)} \cr
&{\tilde F}_4 = {\tilde F}_{5(x)} 
- {\cal B}_{(x)} \wedge \left( {\tilde F}_3  
- j_{(x)} \wedge {\tilde F}_{3(x)} \right)
}}

In the formulae above, a field strength with and without a leg along the $x$ direction are 
 denoted as $F_{n(x)}$, and $F_n$ respectively. 

The inverse of these expressions is given by
\eqn\toremove{\eqalign{
& j_{xx} = {1\over g_{xx}} \cr
& j_{x\alpha}  = - {B_{x\alpha}\over g_{xx}} \cr
& j_{\alpha\beta} = g_{\alpha\beta} 
                   - {1\over g_{xx}} (g_{x\alpha}g_{x\beta}
                   - B_{x\alpha}B_{x\beta})\cr
& {\cal B}_{x\alpha} = - {g_{x\alpha}\over g_{xx}} \cr
& {\cal B}_{\alpha\beta} = { B}_{\alpha\beta} 
                - {1\over g_{xx}} (g_{x\alpha} {B}_{x\beta}
                - {B}_{x\alpha} g_{x\beta}) \cr
& g_s^{IIB} = {g_s^{IIA}\over{\sqrt{g_{xx}}}}
}}
\eqn\twoba{\eqalign{
& {\cal H}_{3(x)}  = - \omega_{(x)} \cr
& {\cal H}_3 = H_3  + \omega_{(x)} \wedge B_{(x)} 
                    - g_{(x)} \wedge H_{3(x)} \cr
}}
and 
\eqn\invforf{\eqalign{
& F_1 = F_{2(x)} \cr
& {\tilde F}_{3(x)}  = F_2 - g_x \wedge F_{2(x)} \cr
& {\tilde F}_3 = {\tilde F}_{4(x)} 
                 - B_{(x)} \wedge \left( F_2 - g_x \wedge F_{2(x)} \right) \cr
& {\tilde F}_{5(x)} = {\tilde F}_{4} - g_{(x)} \wedge {\tilde F}_{4(x)} \cr
}}

The type I theory is equivalent to the heterotic theory by a $S$ duality  
under which the fields are related as 
\eqn\sduality{\eqalign{
& g_s^{\rm het} = {1\over g_s^I} \cr
& j^h_{\mu\nu} = {1\over g_s^I} {\hat j}_{\mu\nu} \cr
& H_3 = {\hat F}_3 \ .
}}
Here we denoted the metric in heterotic theory as $j^h_{\mu\nu}$ and the 
NS field strength $H_3 = dB_2$. $g_s^{\rm het}$ heterotic string coupling.  
Similarly, we denoted the type I metric and RR field strength with 
a ` $\hat{}$ ' where as the string coupling is denoted by a superscript $I$.

\listrefs
\bye